\documentclass{emulateapj}
\usepackage[dvips]{color} 

\newcommand{\etal}{{et al.\@}}

\newcommand{\der}{{\mathrm d}}

\newcommand{\ab}{\alpha_{_{\rm B}}}
\newcommand{\cblue}{}

\def\lsim{\mathrel{\rlap{\lower 4pt \hbox{\hskip 1pt $\sim$}}\raise 1pt\hbox {$<$}}}
\def\gsim{\mathrel{\rlap{\lower 4pt \hbox{\hskip 1pt $\sim$}}\raise 1pt\hbox {$>$}}}

\shorttitle{Spectral evolution of superluminal components}
\shortauthors{Mimica et al.}

\begin{document}

\title{Spectral evolution of superluminal components in parsec-scale
  jets}

\author{P. Mimica\altaffilmark{1}, M.-A.\ Aloy\altaffilmark{1},
  I. Agudo\altaffilmark{2}, J.~M. Mart\'{\i}\altaffilmark{1},  J.~L. G\'omez\altaffilmark{2}}
\author{J.~A. Miralles\altaffilmark{3}} 
\altaffiltext{1}{Departamento de Astronom\'{\i}a y Astrof\'{\i}sica,
  Universidad de Valencia, Edificio de Investigaci\'on, Dr. Moliner
  50, 46100 Burjassot, Spain}
\altaffiltext{2}{Instituto de Astrof\'{\i}sica de Andaluc\'{\i}a
  (CSIC), Apartado 3004, 18080 Granada, Spain}
\altaffiltext{3}{Departament de F\'{\i}sica Aplicada,
  Universitat d'Alacant, Apartado Correos 99, 03080 Alacant, Spain}

%
\begin{abstract}
%
  We present numerical simulations of the spectral evolution and
  emission of {\cblue radio components in relativistic jets. We
    compute jet models by means of a relativistic hydrodynamics
    code. We have developed an algorithm (SPEV) for the transport of a
    population of non-thermal electrons including radiative losses.
    For large values of the ratio of gas pressure to magnetic field
    energy density, $\ab \sim 6\times 10^4$, quiescent jet models show
    substantial spectral evolution, with observational consequences
    only above radio frequencies. Larger values of the magnetic field
    ($\ab \sim 6\times 10^2$), such that synchrotron losses are
    moderately important at radio frequencies, present a larger ratio
    of shocked-to-unshocked regions brightness than the models without
    radiative losses, despite the fact that they correspond to the
    same underlying hydrodynamic structure.  We also show that jets
    with a positive photon spectral index result if the lower limit
    $\gamma_{\rm min}$ of the non-thermal particle energy distribution
    is large enough. A temporary increase of the Lorentz factor at the
    jet inlet produces a traveling perturbation that appears in the
    synthetic maps as a superluminal component. We show that trailing
    components can be originated not only in pressure matched jets,
    but also in over-pressured ones, where the existence of
    recollimation shocks does not allow for a direct identification of
    such features as Kelvin-Helmholtz modes, and its observational
    imprint depends on the observing frequency. If the magnetic field
    is large ($\ab \sim 6\times 10^2$), the spectral index in the
    rarefaction trailing the traveling perturbation does not change
    much with respect to the same model without any hydrodynamic
    perturbation. If the synchrotron losses are considered the
    spectral index displays a smaller value than in the corresponding
    region of the quiescent jet model.}
\end{abstract}
\keywords{galaxies: jets —-- hydrodynamics —-- radiation mechanisms:
  nonthermal —-- relativity}

\maketitle

\section{Introduction}

Relativistic jets are routinely observed emerging from active galactic
nuclei and microquasars, and presumably they are behind the
phenomenology detected in gamma-ray bursts.  It is a broadly
recognized fact that the observed VLBI radio-maps of parsec-scale jets
are not a direct map of the physical state (density, pressure,
velocity, magnetic field) of the emitting plasma. The emission
structure is greatly modified by the fact that a distant (Earth)
observer detects the radiation emitted from a jet which moves at
relativistic speed and forms a certain angle with respect to the line
of sight. Time delays between different emitting regions, Doppler
boosting and light aberration shape decisively the observed aspect of
every time-dependent process in the jet. The observed patterns are
also influenced by the travel path of the emitted radiation towards
the observer since Faraday rotation and, most importantly, opacity
modulate total intensity and polarization radio maps. Finally, there
are other effects that can be very important for shaping VLBI
observations which do not unambiguously depend on the hydrodynamic jet
structure, namely, radiative losses, particle acceleration at shocks,
pair formation, etc. In this work we try to account for some of these
elements by means of numerical simulations.

The basis for currently accepted interpretation of the phenomenology
of relativistic jets was set by \cite{BK79} and \cite{Koenigl81}.  A
number of analytic works have settled the basic understanding that
accounts for the non-thermal synchrotron and inverse Compton emission
of extragalactic jets (e.g., \citealp{Marscher80}), as well as the
spectral evolution of superluminal components in parsec-scale jets
(e.g., \citealp{BM76,HAA85,MGT92,MG85}).  Assuming kinematic jet
models, the numerical implementation of these analytic results enables
one to extensively test the most critical theoretical assumptions by
comparison with the observed phenomenology {\cblue both for steady
  (e.g., \citealp{DM88,HAA89a,HAA91,GAM93,GAM94,Gomezetal94}) and
  unsteady jets (e.g., \citealp{Jones88})}. Basically, the
aforementioned numerical implementation consists on integrating the
synchrotron transfer equations assuming that radiation originated from
an idealized jet model and accounting for all the effects mentioned in
the previous paragraph.

The advent of multidimensional relativistic (magneto-)hydrodynamic
numerical codes has allowed to replace the previously used kinematic,
steady jet models by multidimensional, time-dependent hydrodynamic
models of thermal plasmas (for a review see, e.g.,
\citealp{Gomez02}). The works of \cite{GMMIM95,GMMIA97}, (hereafter
G95 and G97, respectively) \cite{DHO96} or \cite{KF96} compute the
synchrotron emission of relativistic hydrodynamic jet models with
suitable algorithms that account for a number of relativistic effects
(e.g., Doppler boosting, light aberration, time delays, etc.). Their
models assume that there exists a proportionality between the number
and the energy density of the hydrodynamic (thermal) plasma and the
corresponding number and energy density of the emitting population of
non-thermal or supra-thermal particles. These authors assumed that the
magnetic field was dynamically negligible, that the emitted radiation
had no back-reaction on the dynamics, and that that synchrotron losses
were negligible. All these assumptions are very reasonable for VLBI
jets at radio observing frequencies if the jet magnetic field is
sufficiently weak. Consistent with their assumptions, the former
papers included neither a consistent spectral evolution of the
non-thermal particle (NTP) population, nor the proper particle and
energy transport along the jet.

{\cblue The} spectral evolution of NTPs and its transport in classical
jets and radiogalaxies have been carried out by \cite{JRE99},
\cite{Miconoetal99} and \cite{TJR01}. In these works a coupled
evolution of a non-relativistic plasma along with a population of NTPs
has been used to asses either the signatures of diffusive shock
acceleration in radio galaxies \citep{JRE99,TJR01} or the
observational imprint of the non-linear saturation of Kelvin-Helmholtz
(KH) modes developed by a perturbed beam
\citep{Miconoetal99}. \cite{CM03} have also developed a scheme to
perform multidimensional Newtonian magneto-hydrodynamical simulations
coupled with stochastic differential equations adapted to test
particle acceleration and transport in kilo-parsec scale jets.  Dealing
with the spectral evolution of NTPs is relevant in view of the
multiband observations of extragalactic jets where, a significant
aging of the emitting particles seems to be present at optical to
X-ray frequencies (M87, \citealp{HB97,Marshalletal02}; Cen~A,
\citealp{KFJM01}).

This paper builds upon the lines opened by G95 and G97. G95
concentrated on the emission properties from steady relativistic jets,
focusing on the role played by the external medium in determining the
jet opening angle and presence of standing shocks. G97 used a similar
numerical procedure to study the ejection, structure, and evolution of
superluminal components through variations in the ejection velocity at
the jet inlet. \cite{Agudoetal01} discussed in detail how a single
hydrodynamic perturbation triggers pinch body modes in a relativistic,
axisymmetric beam which result in observable superluminal features
trailing the main superluminal component. Finally, \cite{AMGAMI03}
extended the work of \cite{Agudoetal01} to three-dimensional,
helically perturbed beams. Here, we combine multidimensional
relativistic models of compact jets with a new algorithm to compute
the spectral evolution of supra-thermal particles evolving in its
bosom, i.e., including their radiative losses, and their relevance for
the emission and the spectral study of relativistic jets.

{\cblue This work is composed of two parts. In the first part, we
  present a new numerical scheme to evolve populations of relativistic
  electrons in relativistic hydrodynamical flows including radiative
  losses (\S~\ref{sec:SPEV}). For the purpose of calibration the new
  method, our work is} based upon the same axisymmetric, relativistic,
hydrodynamic jet models as employed in G97. Using the same jet
parameters allows us to quantify the relevance of including radiative
losses and, along the way, to compare the emission properties of
parsec-scale jets computed according to two different methods: (1) the
new method presented in this paper and (2) the method presented in G95
and G97, to which we will refer, for simplicity, as {\it Adiabatic
  Method} (AM). {\cblue In the second part of the paper, we apply the
  new method to quantify the relevance of radiative losses in the
  evolution of both quiescent and dynamical jet models.}  We will show
(\S~\ref{sec:radio}) the regimes in which both approaches yield
similar synthetic total intensity radio maps and when synchrotron
losses modify substantially the results. We also show which are the
key parameters to trigger a substantial NTP aging and, therefore, to
significantly change the appearance of the radio maps corresponding to
the same underlying, quiescent jet models. The spectral evolution of a
hydrodynamic perturbation {\cblue travelling} downstream the jet, will
be discussed in Sect.~\ref{sec:spec_evol_component}. Finally, we
{\cblue discuss our main results and conclusions} in
Sect.~\ref{sec:conclusions}.

\section{Hydrodynamic models}
\label{sec:jets}

\begin{table}
\begin{center}
  \caption{{\cblue Set of models used in this work. Values in the table
      refer to the jet nozzle. The first column lists the model
      names. The second and third columns give the
      jet-to-external-medium pressure ratio, and the comoving magnetic
      field at the jet nozzle, respectively. The last column lists the
      values of the ratio of gas pressure to magnetic
      pressure. Additional models not including radiative losses, but
      computed with the SPEV method, will be denoted with a suffix
      ``-NL''. Likewise, models with the same parameters as the ones
      listed here, but computed with the AM method, will be denoted
      with a suffix ``-AM''.}
    \label{tab:models}}
\begin{tabular}{cccc}
  \tableline\tableline
model & $P_b/P_a$ & $b_b$\,[G] & $\ab$            \\ \tableline
PM-S  & 1.0       & 0.002      & $6\times 10^6$ \\
PM-L  & 1.0       & 0.02       & $6\times 10^4$ \\
PM-H  & 1.0       & 0.20       & $6\times 10^2$ \\
OP-L  & 1.5       & 0.03       & $6\times 10^4$ \\
OP-H  & 1.5       & 0.30       & $6\times 10^2$ \\ \tableline
\end{tabular}
\end{center}
\end{table}

Two quiescent, relativistic, axisymmetric jet models constitute our
basic hydrodynamic set up {\cblue (see Tab.~\ref{tab:models})}. They
correspond to the same pressure-matched (PM), and over-pressured (OP)
models of G97. The models were computed in cylindrical symmetry with
the code RGENESIS \citep{MAMB04}. The computational domain spans
$(10R_b\times 200R_b)$ in the $(r \times z)$-plane ($R_b$ is the beam
cross-sectional radius at the injection position). A uniform
resolution of 8 numerical cells/$R_b$ is used. The code module that
integrates the relativistic hydrodynamics equations is a conservative,
Eulerian implementation of a Godunov-type scheme with high-order
spatial and temporal accuracy (based on the GENESIS code;
\citealp{AIMM99,API99}). We follow the same nomenclature as G97 where
quantities affected by subscripts {\it a}, {\it b} and {\it p} refer
to variables of the atmosphere, of the beam at the injection nozzle
and of the perturbation (\S~\ref{sec:perturb}), respectively. The jet
material is represented by a diffuse ($\rho_b / \rho_a=10^{-3}$;
$\rho$ being the rest-mass density), relativistic (Lorentz factor
$\Gamma_b=4$) ideal gas of adiabatic exponent $4/3$, with a Mach
number $M_b =1.69$. At the injection position, model PM has a pressure
$P_b = P_a$, while model OP has $P_b = 1.5P_a$.%
%
%
%
 Pressure in the atmosphere decays with distance $z$ according to
$P(z) = P_a/[1 + (z/z_c)^{1.5}]^{1.53}$, where $z_c = 60R_b$.
%
%
With such an atmospheric profile both jet models display a paraboloid
shape, which introduces a small, distance-dependent, jet opening angle
which is compatible with observations of parse-scale jets.  At a
distance of $200R_b$, the opening angles for the models PM and OP are
$0.29^\circ$ and $0.43^\circ$, respectively.

Pressure equilibrium in the atmosphere is ensured by including
adequate counter-balancing, numerical source terms. However, despite
the fact that the initial model is very close to equilibrium, small
numerical imbalance of forces triggers a transient evolution that
decays into a final quasi-steady state after roughly $2-5$
longitudinal grid light-crossing times. {\cblue We} treat these
quiescent states as initial models. Model PM yields an {\cblue
  adiabatically-expanding, smooth beam}. Model OP develops a
collection of cross shocks in the beam, {\cblue whose spacing
  increases with the distance from the jet basis}.

\subsection{Injection and Evolution of Hydrodynamic Perturbations}
\label{sec:perturb}

Variations in the injection velocity (Lorentz factor) have been
suggested as a way to generate internal shocks in relativistic jets
\citep{Rees78}. We set up a traveling perturbation in the jet as a
sudden increase of the Lorentz factor at the jet nozzle (from
$\Gamma_b=4$ to $\Gamma_p=10$) for a short period of time
($0.75R_b/c$; $c$ being the light speed). Since the injected
perturbation is the same as in G97, its evolution is identical to the
one these authors showed and, thus, we provide a brief overview
here. The perturbation develops two Riemann fans emerging from its
leading and rear edges (see, e.g., \citealp{MAMB05,MAM07}). In front
of the perturbation a shock-contact discontinuity-shock structure
(${\cal SCS}$) forms, while the rear edge is trailed by a
rarefaction-contact discontinuity-rarefaction (${\cal RCR}$) fan. In
the leading shocked region the beam expands radially owed to the
pressure increase with respect to the atmosphere. In the trailing
rarefied volume the beam shrinks radially on account of the smaller
pressure in the beam than in the external medium. This excites the
generation of pinch body modes in the beam that seem to trail the main
hydrodynamic perturbation as pointed out by \cite{Agudoetal01}. Also
the component itself splits in, at least, two parts when the forward
moving rarefaction leaving the rear edge of the component merges with
the reverse shock traveling backwards (in the component rest frame)
that leaves from the forward edge of the hydrodynamic perturbation
({\cblue as in} \citealp{AMGAMI03}).


\section{SPEV: A new algorithm to follow non-thermal particle
  evolution}
\label{sec:SPEV}

The {\it sp}ectral {\it ev}olution (SPEV) routines are a set of
methods developed to follow the evolution of NTPs in the phase
space. {\cblue Here} we assume that the radiative losses at radio
frequencies are negligible with respect to the total thermal energy of
the jet at every point in the jet. Thus, radiation back reaction onto
the hydrodynamic evolution is neglected. Certainly, such an {\it
  ansatz} is invalid at shorter wavelengths (optical, X-rays), where
radiative losses shape the observed spectra (see, e.g., \cite{MAMB05}
for X-ray-synchrotron blazar models that include the radiation
back-reaction onto the component dynamics).

The {\cblue 7-}dimensional space formed by the particle momenta,
particle positions and time is split into two parts.
%
%
For the spatial part of the phase space, we assume that NTPs do not
diffuse in the hydrodynamic (thermal) plasma. Thereby, the spatial
evolution of the NTPs is governed by the velocity field of the
underlying fluid, and it implies that the NTP comoving frame is the
same as the thermal fluid comoving frame. Assuming a negligible
diffusion of NTPs is a sound approximation in most parts of our
hydrodynamic models since the electron diffusion lengths are much
smaller than the dynamical lengths in smooth flows (see, e.g.,
\citealp{TJR01,Miniati01}). Obviously, the assumption is not fulfilled
wherever diffusive acceleration of NTPs takes place (e.g., at shocks
or at the jet lateral boundaries). Nevertheless, there exists a strong
mismatch between the scales relevant to dynamical and diffusive
transport processes for NTPs of relevance to synchrotron
radio-to-X-ray emissions within relativistic jets. The mismatch
ensures that even in macroscopic, non-smooth regions such as the cross
shocks in the beam of model OP, the assumption we have made suffices
to provide a good qualitative description of the NTP population
dynamics.

Consistent with the hydrodynamic discretization, we assume that the
velocity field is uniform inside each numerical cell (equal to the
average of the velocity inside such cell). In practice, a number of
Lagrangian {\it particles} are introduced through the jet nozzle, each
evolving the same NTP distribution but being spatially transported
according to the local fluid conditions.  We emphasize that these
Lagrangian particles are used here for the solely purpose of
representing the spatial evolution (i.e., the trajectories) of
ensembles of NTPs. We integrate the trajectories of such particles
using a conventional time-explicit, adaptive-step-size, fourth order
Runge-Kutta (RK) integrator.

\subsection{Particle Evolution in the momentum space}

In order to derive the equations governing the time evolution of
charged NTPs in the momentum space we follow closely the approach
of \cite{MRL93} (see also \citealp{Webb85}, or \citealp{Kirk94}). We
start by considering the Boltzmann equation that obeys the ensemble
averaged distribution function $f$ of the NTPs, each with a rest-mass
$m_0$,
\begin{equation}
p^\beta\left(\frac{\partial f}{\partial x^\beta}-
\Gamma^\alpha_{\beta\gamma}p^\gamma\frac{\partial f} {\partial p^\alpha}\right)
=\left(\frac{df}{d\tau}\right)_{coll},
\label{eq:Boltzman}
\end{equation}
where $f$ is a function of the coordinates $x^\alpha$ and the components of 
the particle 4-momentum $p^\alpha$ with respect to the coordinate basis 
${\bf e}_{(\alpha)}$. The $\Gamma^\alpha_{\beta\gamma}$ are the usual 
Christoffel symbols and the right hand side represents the collision term, 
$\tau$ being the particle proper time. 

{\cblue Equation~\ref{eq:Boltzman}} can be written in terms of the
particle 4-momentum components with respect to the comoving or matter
frame instead of the components with respect to the coordinate
basis. The comoving tetrad ${\bf e}_{(a)}$ ($a=0, 1, 2, 3$), is formed
by four vectors, one of which (${\bf e}_{(0)}$) is the four velocity
of the matter and the following orthonormality relation is fulfilled
\begin{eqnarray*}
{\bf e}_{(a)}\cdot{\bf e}_{(b)}=\eta_{ab},
\end{eqnarray*}
where $\eta_{ab}$ is the Minkowski metric ($\eta_{00}=-1$). We
explicitly point out that the components of tensor quantities with
respect to the coordinate and tetrad basis are annotated with Greek and
Latin indices, respectively.  The transformation between the basis
${\bf e}_{(\alpha)}$ and ${\bf e}_{(a)}$ is given by the matrix
$e^\alpha_a$ and its inverse matrix $e'^a_\alpha$,
\begin{equation}
{\bf e}_{(a)}=e^\alpha_a {\bf e}_{(\alpha)}, \;\; {\bf e}_{(\alpha)}=e'^a_\alpha
{\bf e}_{(a)}
\end{equation}

 In terms of the comoving basis, the Boltzmann equation is
\begin{equation}
p^b\left(e^\beta_b\frac{\partial f}{\partial x^\beta}-
\Gamma^a_{bc}p^c\frac{\partial f} {\partial p^a}\right)
=\left(\frac{\delta f}{\delta\tau}\right)_{\rm coll}.
\label{eq:Boltzmann}
\end{equation}
{\cblue The connection coefficients in the tetrad frame
  $\Gamma^a_{bc}$ obey the following relations}
\begin{equation}
\Gamma^a_{bc}=e^\beta_b e'^a_\alpha e^\alpha_{c;\beta}=e^\beta_b
e'^a_\alpha\left(e^\alpha_{c,\beta}+\Gamma^\alpha_{\beta\gamma}
e^\gamma_c\right),
\end{equation}
where the comma {\cblue and the semicolon stand for partial and
  covariant derivatives, respectively}.

We introduce the two first moments of the distribution function by the
equations
\begin{equation}
n^a=\int d\Omega \frac{\rm p^2}{p^0}p^af,
\label{eq:n^a}
\end{equation}
\begin{equation}
t^{ab}=\int d\Omega \frac{\rm p^2}{p^0}p^ap^bf,
\end{equation}
where p$^2=(p^0)^2-m_0^2c^2$ is the square of the NTP three-momentum
measured by the comoving observer. The solid-angle ($\Omega$)
integrations are performed over all particle momentum
directions. {\cblue The number of NTPs per unit volume with modulus of
  their three-momentum between p and p+$d{\rm p}$ for an observer
  comoving with the matter is $n^0({\rm p})d{\rm p}$}. Further
integration of the above moments $n^a$ and $t^{ab}$ over p,
$\int_0^\infty d{\rm p}$, {\cblue gives} the hydrodynamical moments.

In order to obtain the continuity equation for NTPs, we multiply the
Boltzmann equation~(\ref{eq:Boltzmann}) by $({\rm p}^2 / p^0)$, and
integrate over $\Omega$ to yield (for the details see {\cblue App.~A
  of} \citealp{Webb85}),

\begin{equation}
e^{\alpha}_a \frac{\partial n^a}{\partial x^\alpha} + 
e^{\alpha}_{a; \alpha} n^a - 
\frac{\partial}{\partial {\rm p}} 
   \left( \frac{p^0}{\rm p} \Gamma^0_{ab} t^{ab} \right) = 
\int d\Omega\, \frac{{\rm
    p}^2}{p^0} \left( \frac{\delta f}{\delta \tau}\right)_{\rm coll}\, .
\label{eq:cont}
\end{equation}

The next step is to formulate the continuity equation in the diffusion
approximation. Such approximation implies that the scattering of NTPs
by hydromagnetic turbulence results in a quasi isotropic distribution
function in the scattering (comoving) frame. Thus, it is assumed that
the distribution function of the NTPs can be expressed as the sum of
two terms, $f=f^{(0)}+f^{(1)}{\bf \Omega}$, where $f^{(0)}\gg f^{(1)}$
and ${\bf \Omega}$ is the unit vector in the direction of the momentum
of the particle. With such an assumption, we obtain that
\begin{equation}
  n^0 \simeq 4\pi {\rm p}^2 f^{(0)} \gg n^i\, , \:\,\,\, t^{ij} \simeq \frac{{\rm
      p}^2}{p^0}\frac{\delta^{ij}}{3} n^0\, , \:\,\,\, i,j=1,2,3\, ,
\label{eq:diffusion-approx1}
\end{equation}
which also leads to
\begin{equation}
t^{00}\simeq p^0 n^0\gg t^{0i}=t^{i0}.
\label{eq:diffusion-approx2}
\end{equation}

Plugging the approximations (\ref{eq:diffusion-approx1}) and
(\ref{eq:diffusion-approx2}) into Eq.~(\ref{eq:cont}) and neglecting
the terms coming from the anisotropy of the distribution function,
i.e., the terms arising from $f^{(1)}$, we obtain
\begin{equation}
e^{\alpha}_0 \frac{\partial n^0}{\partial x^\alpha} + 
e^{\alpha}_{0; \alpha} n^0 - 
\frac{\partial}{\partial {\rm p}} 
   \left( \sum_{i=1}^3 \Gamma^i_{0i} {\rm p} \frac{n^0}{3} \right) = 
\int d\Omega\, \frac{{\rm
    p}^2}{p^0} \left( \frac{\delta f}{\delta \tau}\right)_{\rm coll}\, .
\label{eq:cont-diff_approx}
\end{equation}

Equation~(\ref{eq:cont-diff_approx}) is valid for any general metric
$g_{\mu \nu}$. However, in the present work we are only interested in
obtaining the transport equation for NTPs in the special relativistic
regime. To restrict Eq.~(\ref{eq:cont-diff_approx}) to such a regime
we take a flat metric, $g_{\mu\nu} = \eta_{\mu \nu}$. {\cblue
  Thereby,} the tetrad and the coordinate frame basis are related by a
simple Lorentz transformation, i.e.,
\begin{eqnarray*}
e^{0}_0 &=& \Gamma,\, \\
e^{0}_i &=& e^{i}_0 = \Gamma v^i, \, \\
e^{i}_j &=& \delta_{ij} + (\Gamma -1) \frac{v^i v^j}{v^2} \:\:\: (i,j=1,2,3),
\label{eq:lorentz_transformation}
\end{eqnarray*}
where $v^i$ $(i=1,2,3)$ are the spatial components of the velocity of
matter, which is equal to the hydrodynamical velocity of the NTPs,
since we make the assumption that NTPs do not diffuse in the
hydrodynamic plasma. The hydrodynamical Lorentz factor of the plasma
is denoted by $\Gamma=1/\sqrt{(1-v^iv_i)}$.  With this transformation we
obtain
\begin{eqnarray}
e^{\alpha}_{0; \alpha} &=& e^{\alpha}_{0, \alpha} =
\frac{\partial \Gamma}{\partial t} + \frac{\partial \Gamma
  v^i}{\partial x^i} = \Theta 
\label{eq:lorentz_transformation2} \\
\sum_{i=1}^3 \Gamma^i_{0i} &=& 
\frac{\partial \Gamma}{\partial t} + \frac{\partial \Gamma
  v^i}{\partial x^i} = \Theta \, ,
\label{eq:lorentz_transformation3}
\end{eqnarray}
$\Theta$ being the expansion of the underlying thermal fluid, which is
related to {\cblue $\rho$} by
\begin{eqnarray}
  \Theta = -\frac{D \ln{\rho}}{D\tau}\, .
\label{eq:expansion}
\end{eqnarray}

Plugging Eqs.~(\ref{eq:lorentz_transformation2})-(\ref{eq:expansion})  into
Eq.~(\ref{eq:cont-diff_approx}) and using the definition of the
Lagrangian derivative with respect to the proper time of the comoving observer
\begin{eqnarray*}
  \frac{D}{D\tau} = \Gamma \left( 
\frac{\partial}{\partial t} + v^i \frac{\partial}{\partial x^i}
\right)\, ,
\end{eqnarray*}
 yields
\begin{equation}
  \frac{D n^0}{D \tau} + \Theta n^0 -
  \frac{\partial}{\partial {\rm p}}
  \left( \frac{n^0}{3} {\rm p} \Theta \right) = 
  \int d\Omega\, \frac{{\rm
      p}^2}{p^0} \left( \frac{\delta f}{\delta \tau}\right)_{\rm coll}\, ,
\label{eq:cont-diff_approx_SR_0}
\end{equation}
which can be cast in the form
\begin{equation}
  \frac{D \ln{n^0}}{D \tau}  - 
  \frac{{\rm p}}{3} \Theta \frac{\partial
    \ln{n^0}}{\partial {\rm p}} 
+ \frac{2}{3}\Theta
= \frac{1}{n^0}
  \int d\Omega\, \frac{{\rm
      p}^2}{p^0} \left( \frac{\delta f}{\delta \tau}\right)_{\rm coll}\, .
\label{eq:cont-diff_approx_SR_1}
\end{equation}

The collision term contains the interaction between NTPs and matter,
radiative losses due to synchrotron processes, etc. Let us consider
first the interaction with matter. In this case, the collisions can be
assumed to be isotropic in the comoving frame and elastic. In such a
case, and consistently to the previous approximation, the collision
term in Eq.~(\ref{eq:cont-diff_approx_SR_1}) vanishes and we can find
a solution for the homogeneous differential equation by considering
\begin{eqnarray*}
  \frac{D \ln{n^0}}{D \tau}  - 
  \frac{{\rm p}}{3} \Theta \frac{\partial
    \ln{n^0}}{\partial {\rm p}} \, ,
\end{eqnarray*}
as the derivative of $\ln{n^0}$ along the following curve in the plane 
($\tau$,p), parametrized by $\sigma$
\begin{eqnarray*}
  \frac{d \tau}{d\sigma} & = &1 \\
  \frac{d {\rm p}}{d \sigma} &=&  -\frac{{\rm p}}{3} \Theta\, ,
\end{eqnarray*}
i.e., we may write Eq.~(\ref{eq:cont-diff_approx_SR_1}) as
\begin{eqnarray}
\frac{d \ln{n^0}}{d\sigma} & =& -\frac{2}{3} \Theta\, .
\label{eq:deriv2}
\end{eqnarray}
The solution of Eq.~(\ref{eq:deriv2}), is
\begin{eqnarray}
n^0(\tau(\sigma), p(\sigma))  = n^0(\tau(\sigma_0), p(\sigma_0)) 
\left( \frac{\rho(\tau(\sigma))}{\rho(\tau(\sigma_0))} \right)^{\frac{2}{3}}\, ,
\label{eq:n^0_homogeneous}
\end{eqnarray}
where $\sigma_0$ corresponds to some initial value of the parameter
$\sigma$. Equation~(\ref{eq:n^0_homogeneous}) expresses the fact that
the variation of the number of NTPs per unit of volume along a certain
curve (parametrized by $\sigma$) is directly related with the
variation of the rest-mass density of the thermal plasma
between the initial and final points of such a curve.

We now turn back to Eq.~(\ref{eq:cont-diff_approx_SR_1}) and derive
the form of the collisions term in the case that the only relevant
radiative losses are due to synchrotron processes. In such a case we
have \citep[e.g.,][]{RL79}
\begin{equation}
  \left( \frac{d{\rm p}}{d\tau} \right)_{\rm syn} =
  - \frac{4 \sigma_{\rm T} {\rm p}^2 U_{\rm B}}{3 m_{\rm e}^2 c^2}
  := {\cal B} ({\rm p},\tau)\, ,
\label{eq:dpdt_syn}
\end{equation}
where $\sigma_{\rm T}$ is the Thompson cross section, $m_{\rm e}$ is
the electron rest-mass, $U_B={\bf b}^2/8\pi$ is the magnetic energy
density, and we have assumed that the electrons are ultrarelativistic,
$p \approx \gamma m_{\rm e}$, $\gamma$ {\cblue being} the electron
Lorentz factor (not to be confused with the plasma Lorentz
factor). {\cblue If $\Theta=0$, Eq.~(\ref{eq:cont-diff_approx_SR_1})
  reads in the comoving frame}
\begin{equation}
 \left( \frac{D \ln{n^0}}{D\tau} \right)_{\rm syn} = \frac{1}{n^0}
\int d\Omega\, \frac{{\rm
      p}^2}{p^0} \left( \frac{\delta f}{\delta \tau}\right)_{\rm coll}
\label{eq:dndtau_syn}
\end{equation}
and, on the other hand, the particle number conservation yields
\begin{eqnarray*}
 \left( \frac{Dn^0}{D\tau} \right)_{\rm syn} = -
 \frac{\partial }{\partial {\rm p}} \left( n^0 {\cal B}\right)\, ,
\end{eqnarray*}
or, equivalently,
\begin{equation}
 \left( \frac{D \ln{n^0}}{D\tau} \right)_{\rm syn} = 
    - \frac{\partial {\cal B}}{\partial {\rm p}}
    - {\cal B} \frac{\partial \ln{n^0}}{\partial {\rm p}}\, .
\label{eq:dndtau_syn2}
\end{equation}

Taking into account Eq.~(\ref{eq:dndtau_syn}), we may plug
Eq.~(\ref{eq:dndtau_syn2}) into  Eq.~(\ref{eq:cont-diff_approx_SR_1})
to account for the combined effects of synchrotron losses and
adiabatic expansion/compression of the fluid
\begin{equation}
  \frac{D \ln{n^0}}{D \tau}  + 
  \left( -\frac{{\rm p}}{3} \Theta + 
     {\cal B} \right) \frac{\partial \ln{n^0}}{\partial {\rm p}} 
= - \frac{2}{3}\Theta - 
  \frac{\partial {\cal B}}{\partial {\rm p}}\, .
\label{eq:adiabatic_syn}
\end{equation}
The formal solution of Eq.~(\ref{eq:adiabatic_syn}), can be found
following the same procedure we used above for the homogeneous
continuity equation. In this case we interpret
\begin{eqnarray*}
  \frac{D \ln{n^0}}{D \tau}  + 
  \left( -\frac{{\rm p}}{3} \Theta + 
    {\cal B} \right) \frac{\partial \ln{n^0}}{\partial {\rm
      p}}\, ,
\end{eqnarray*}
as the derivative of $\ln{n^0}$ along the curve
\begin{eqnarray}
  \frac{d \tau}{d\sigma} & = &1 \nonumber \\
  \frac{d {\rm p}}{d \sigma} &=&  -\frac{{\rm p}}{3} \Theta + 
  {\cal B}\, ,
\label{eq:dpdsigma}
\end{eqnarray}
which yields, on the one hand,
\begin{eqnarray}
  \frac{d {\rm p}}{d \tau} &=&  -\frac{{\rm p}}{3} \Theta + 
{\cal B}({\rm p},\tau)\, ,
\label{eq:dpdt}
\end{eqnarray}
and, on the other hand,
\begin{eqnarray}
  n^0(\tau(\sigma), p(\sigma)) & =& n^0(\tau(\sigma_0), p(\sigma_0)) 
  \left(
    \frac{\rho(\tau(\sigma))}{\rho(\tau(\sigma_0))}\right)^{\frac{2}{3}}
  \nonumber \\
  & & \times \exp{\left(-\int_{\sigma_0}^{\sigma} d\sigma' \frac{\partial
      {\cal B}({\rm p},\tau)}{\partial {\rm p}}(\sigma')\right)}\, .
\label{eq:n^0}
\end{eqnarray}
Equation~(\ref{eq:dpdt}) shows the evolution of the particle momentum
in time, while Eq.~(\ref{eq:n^0}) is only a formal solution since the
exact dependence of ${\rm p}(\sigma)$, necessary to perform the
integration, is only known through the differential
equation~(\ref{eq:dpdsigma}). The first term on the right hand side of
Eq.~(\ref{eq:dpdt}) accounts for the change of momentum due to the
adiabatic expansion or compression of the fluid in which NTPs are
embedded.  The time dependence of ${\cal B}$ is fixed by the
hydrodynamic properties of the thermal fluid and by the comoving
magnetic field ${\bf b}$, assumed to be provided by hydrodynamic
simulations and models of the ${\bf b}$-field (which is not directly
simulated), respectively.

In order to speed up the numerical evaluation of Eqs.~(\ref{eq:dpdt})
and (\ref{eq:n^0}), we assume that both, the fluid expansion and the
synchrotron losses (or, equivalently, {\cblue $U_{\rm B}$}) are
constant within an small interval of proper time around
$\tau(\sigma_0)$. Thus, we can write Eq.~(\ref{eq:dpdsigma}) as
\begin{eqnarray}
  \frac{d {\rm p}}{d \sigma} &=& k_{\rm a} {\rm p} - k_{\rm s} {\rm
    p}^2\, ,
\label{eq:dpdsigma2}
\end{eqnarray}
$k_{\rm a}$ and $k_{\rm s}$ being both
constants, such that the following relations hold
\begin{eqnarray}
 \frac{\rho(\tau(\sigma))}{\rho(\tau(\sigma_0))} &=& {\rm e}{^{3k_{\rm
     a}\Delta\sigma}} \\ \label{eq:ka_ks}
  {\cal B}({\rm p}(\sigma),\tau(\sigma)) &=& -k_{\rm s}{\rm p}^2\, ,
\end{eqnarray}
with $\Delta\sigma=\sigma - \sigma_0$. Equation~(\ref{eq:dpdsigma2})
has the following analytic solution
\begin{equation}
  {\rm p}(\sigma) = {\rm p}_0 \frac{k_{\rm a} {\rm e}^{k_{\rm a}\Delta\sigma}}{k_{\rm a} +  {\rm p}_0k_{\rm s} \left( {\rm e}^{k_{\rm
      a}\Delta\sigma} - 1\right)}\, ,
\label{eq:p(sigma)}
\end{equation}
where ${\rm p}_0 := {\rm p}(\sigma_0)$. Upon substitution of the
relations~(\ref{eq:ka_ks}) and (\ref{eq:p(sigma)}) in
Eq.~(\ref{eq:n^0}) we obtain
\begin{eqnarray}
  n^0(\tau(\sigma),{\rm p}(\sigma)) &=& n^0(\tau(\sigma_0),
  p(\sigma_0)) \times \nonumber \\
  & &  \left[ {\rm e}{^{k_{\rm a}\Delta\sigma}} 
    \left( 1 + {\rm p}_0 \frac{k_{\rm s}}{k_{\rm a}} \left({\rm e}^{k_{\rm a}\Delta\sigma} -1\right) \right) \right]^2\, .
\label{eq:n^0(sigma)}
\end{eqnarray}

This equation is approximately valid in the neighborhood of
$\tau(\sigma_0)$ or if the fluid expansion and magnetic field energy
are both constant in a certain interval $\Delta\sigma$. Indeed, such
an assumption is adequate for our purposes, since the hydrodynamic
evolution is performed numerically as a succession of finite, but
small, time steps. Within each hydrodynamic time step the physical
variables inside of each numerical cell do not change much and, thus,
the magnetic field energy and the fluid expansion are roughly
constant. Alternatively, one might not assume anything about $\Theta$
or $U_{\rm B}$ and solve the system of integro-differential equations
(\ref{eq:dpdsigma}) and (\ref{eq:n^0}). However, such a procedure is
much more computationally demanding than obtaining the evolution of
${\rm p}$ and $n^0$ from, respectively, Eqs.~(\ref{eq:p(sigma)}) and
(\ref{eq:n^0(sigma)}). Furthermore, since the magnetic field is
assumed in this work, i.e., not consistently computed, a numerical
solution of the aforementioned equations does not yield a true
improvement of the accuracy.

For completeness, {\cblue as} in the diffusion approximation
$n^0=4\pi{\rm p}^2f^{(0)}$, we can specify the evolution equation for
the isotropic part of the distribution function of the NTPs
\begin{eqnarray}
 f^{(0)}(\tau(\sigma),{\rm p}(\sigma)) &=& f^{(0)}_0  
  \left( 1 + {\rm p}_0 \frac{k_{\rm s}}{k_{\rm a}} \left({\rm
        e}^{k_{\rm a}\Delta \sigma} -1\right) \right)^4 ,
\label{eq:f(sigma)}
\end{eqnarray}
where $f^{(0)}_0=f^{(0)}(\tau(\sigma_0), p(\sigma_0))$.

Finally, we define the number density of NTPs within a certain
momentum interval $[{\rm p}_a(\tau(\sigma)), {\rm p}_b(\tau(\sigma))]$
\begin{eqnarray}
  {\cal N}(\tau(\sigma);{\rm p}_a,{\rm p}_b) &:=& \int_{{\rm p}_a(\tau(\sigma))}^{{\rm p}_b(\tau(\sigma))} 
   d{\rm p}\, n^0(\tau(\sigma_0), p(\sigma_0))\, ,
\label{eq:n_def}
\end{eqnarray}
whose evolution equation can be easily obtained from
Eqs.~(\ref{eq:p(sigma)}) and (\ref{eq:n^0(sigma)}) and reads
\begin{eqnarray}
  {\cal N}(\tau(\sigma);{\rm p}_a,{\rm p}_b) &=& 
{\rm e}^{3k_{\rm a}\Delta\sigma} {\cal N}(\tau(\sigma_0);{\rm p}_a,{\rm p}_b)\, .
\label{eq:n(tau)}
\end{eqnarray}

Equation(\ref{eq:n(tau)}) shows that the time evolution of the number
density of NTPs in a time-evolving momentum interval, depends only on
the adiabatic changes of the NTPs in such momentum interval, but not
on the synchrotron losses {\cblue (Eq.~(\ref{eq:n(tau)}) is
  independent of $k_{\rm s}$).}

\subsection{Discretization in momentum space}
\label{sec:discretization}

In the following we normalize $p$ to $m_{\rm e} c$, which allows us to
express our results in terms of the particle Lorentz factor $\gamma$
instead of $p$.  In order to make Eqs.~(\ref{eq:p(sigma)}) and
(\ref{eq:n^0(sigma)}) amenable to numerical treatment, we discretize
the momentum space in $N_{\rm b}$ bins, each momentum bin $i$ having a
lower bound $\gamma_i$.  In the present applications we use $N_{\rm b}
= 32$ {\cblue (see App.~\ref{sec:totalflux}). As in, e.g.,
  \cite{JRE99}, \cite{Miniati01}, or \cite{JK05},} we initially
distribute $\gamma_i$ logarithmically, i.e.
\begin{eqnarray*}
\gamma_i(\tau_0) = \gamma_{\rm min}\left( \frac{\gamma_{\rm max}}{\gamma_{\rm min}}
\right)^{(i-1)/(N_{\rm b}-1)}\, ,
\end{eqnarray*}
$\gamma_{\rm min}$ and $\gamma_{\rm max}$ being the minimum and
maximum Lorentz factors of the considered distribution,
respectively. 

On the other hand, the time dimension is also discretized in time
steps. We call $\tau^n$ the interval of proper time elapsed since the
beginning of our simulation, and denote $\Delta \tau =
\tau^{n+1}-\tau^n$.

Our numerical method follows the time evolution of NTPs in the
momentum space employing a Lagrangian approach. We track both the
evolution of the $N_{\rm b}$ interface values $n^0_i$ (from
Eq.~(\ref{eq:n^0(sigma)}), where we take $\tau=\sigma$ and
$\sigma_0=\tau(\sigma_0):=\tau^n$),

\begin{eqnarray}
  n^0_i(\tau^{n+1})&:=&n^0(\tau^{n+1}, \gamma_i(\tau^{n+1})) = n^0(\tau^n, \gamma_i(\tau^n)) \times \nonumber \\
  & &  \left[ {\rm e}{^{k_{\rm a}\Delta\tau}} 
    \left( 1 + \gamma_i(\tau^n) \frac{k_{\rm s}}{k_{\rm a}} \left({\rm
          e}^{k_{\rm a}\Delta \tau} -1\right) \right) \right]^2\, ,
\label{eq:n_i^0}
\end{eqnarray}
as well as the $N_{\rm b}$ bin integrated values
\begin{equation}
{\cal N}_i(\tau) :=\int_{\gamma_i(\tau)}^{\gamma_{i+1}(\tau)} d\gamma\, n^0(\tau, \gamma)\, .
\label{eq:n_i(tau)}
\end{equation}

The time evolution of the $N_{\rm b} + 1$ interface values
$\gamma_i(\tau)$ is governed by Eq.~(\ref{eq:p(sigma)}).

\begin{figure*}
\plotone{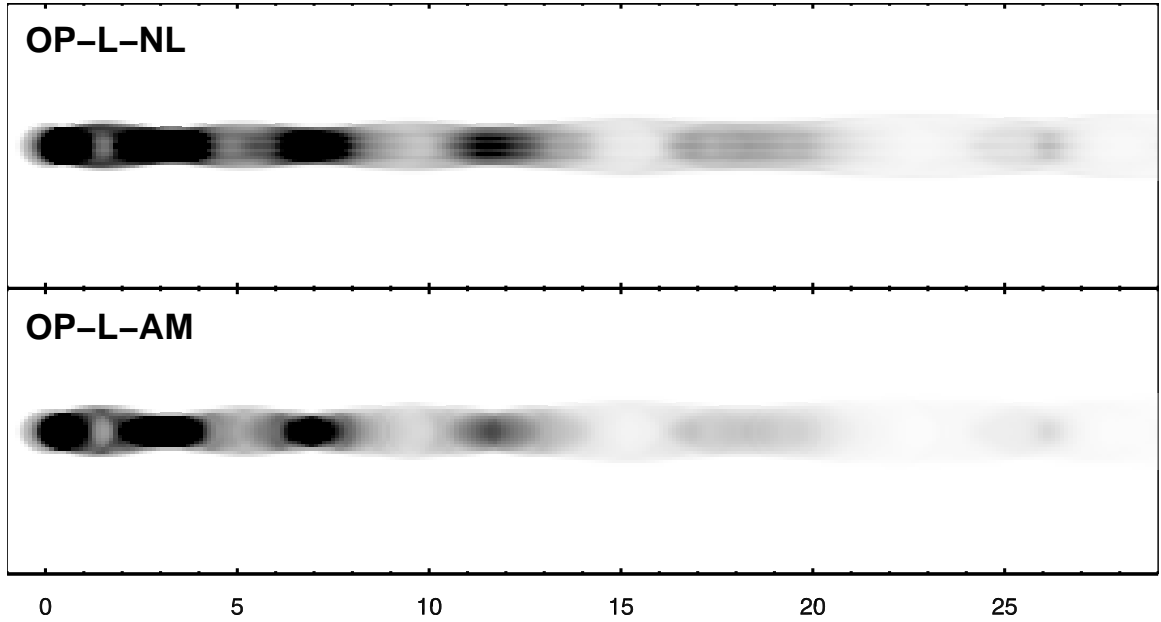}
\caption{Top panel: Synthetic (total intensity) radio map of the
  quiescent {\cblue OP-L-NL model} computed at an observational frequency of
  43~GHz. Bottom panel: Same as top panel but {\cblue for model OP-L-AM},
  and plotted using the same intensity scale. To compute {\cblue the model OP-L-NL}, 32 Lagrangian particles evenly distributed across the
  jet nozzle have been let to evolve. Since SPEV-NL does not include
  the effect of synchrotron losses on the NTP evolution, the
  differences between both radio maps are small. In both panels a
    $10^\circ$ jet viewing angle is assumed.
    \label{fig:fiduc-emiss}}
\end{figure*}

For the purpose of efficiently computing the synchrotron emissivity
(see \S~\ref{sec:synchrotron}), inside of each Lorentz factor bin $i$,
we assume that, at any time, the number of NTPs per unit of energy and
unit of volume $n^0_i(\tau,\gamma)$ ($\gamma_i(\tau) \le \gamma <
\gamma_{i+1}(\tau)$) follows a power-law and, therefore, the whole
momentum distribution of NTPs consists of a piecewise power-law and,
\begin{eqnarray}
  n^0(\tau,\gamma) = n^0_i(\tau) \left(
    \frac{\gamma}{\gamma_i}\right)^{-q_i(\tau)},\:  i=1, \dots, N_{\rm
    b}, 
\label{eq:n^0_powerlaw}
\end{eqnarray}
where $n^0_i(\tau)$ is the number of particles with $\gamma=\gamma_i$
at the proper time $\tau$, and $q_i(\tau)$ is the power-law index of
the distribution at the $i$-Lorentz factor interval.  The values of
$q_i(\tau)$ are computed numerically in every time step plugging
Eq.~(\ref{eq:n^0_powerlaw}) into Eq.~(\ref{eq:n_i(tau)}) and solving
iteratively the corresponding equation (which also involves knowing
the interface values -Eq.~(\ref{eq:n_i^0})- and justifies why we need
to follow the evolution of two sets of variables {\cblue per} bin).

The approach defined up to here has the advantage that, at every time
level $\tau^n$, the momentum-space evolution and the physical space
trajectory of the NTPs are decoupled during the corresponding time
step $\Delta \tau$. The hydrodynamic evolution of the thermal plasma
provides the values of $k_{\rm a}$ and $k_{\rm s}$ at the beginning of
the time step ($\tau=\tau^n$), and once these values are known, it is
possible to compute the momentum distribution of NTPs at time
$\tau^{n+1}$. Thereby, it is possible to perform separately the
trajectory integration of the NTPs once, and to evolve NTPs in the
phase space afterward, as many times and with as many initial
particle distributions as desired (viz., during a post-processing
phase).

\subsection{Normalization and initialization of the NTP distribution}

Our models are set up such that we initially inject through the jet
nozzle NTPs with a momentum distribution function which follows a
single power-law, i.e., $q_i=q_1, \forall i$. Therefore, the initial
number and energy density in the interval $\gamma_{\rm min} \leq
\gamma \leq \gamma_{\rm max}$ read
\begin{eqnarray}
 {\cal N} & =& \frac{n^0_1}{q_1-1}\gamma_{\rm min}
      \left[ 1-\left(\frac{\gamma_{\rm max}}{\gamma_{\rm min}}\right)^{1-q_1}
      \right] \label{eq:N} ,
\\
{\cal U} &= & \frac{n^0_1}{q_1-2} \gamma_{\rm min}^2 m_{\rm e} c^2
      \left[1-\left(\frac{\gamma_{\rm max}}{\gamma_{\rm
              min}}\right)^{2-q_1}\right]\, .
\label{eq:E}
\end{eqnarray}

Consistent with our assumptions about the relation between the thermal
and non-thermal populations we assume that ${\cal N} = c_{_{\cal N}}
\rho/m_{\rm e}$ and ${\cal U} = c_{_{\cal U}} P$, where $c_{_{\cal
    N}}$ and $c_{_{\cal U}}$ are constants, while $P$ and $\rho$ stand
for the pressure and rest-mass density of the background fluid,
respectively. Such proportionalities along with Eqs.~(\ref{eq:N}) and
(\ref{eq:E}) yield (G95)
\begin{equation} 
  \gamma_{\rm min} =
  \frac{c_{_{\cal U}}}{c_{_{\cal N}}}\frac{q_1-2}{q_1-1}\frac{P}{\rho
    c^2}\frac{1-(\gamma_{\rm
      max}/\gamma_{\rm min})^{1-q_1}}{1-(\gamma_{\rm max}/\gamma_{\rm min})^{2-q_1}},
\label{eq:pmin}
\end{equation}
and we can use either Eq.~(\ref{eq:N}) or Eq.~(\ref{eq:E}) to compute
$n^0_1$ if the ratio $C_\gamma:=\gamma_{\rm max}/ \gamma_{\rm min}$ is
fixed. Thus, the initial distribution of particles can be determined
from pressure and rest-mass density at the jet nozzle, simply by
specifying $c_{_{\cal N}}$ and $c_{_{\cal U}}$.

A key difference between SPEV and AM methods is that in SPEV the
dimensionless proportionality parameters $c_{_{\cal N}}$ and
$c_{_{\cal U}}$ are only specified at the jet injection nozzle. In the
SPEV method, the subsequent time evolution of the NTP momentum
distribution, namely, the spectral shape (piecewise power-law) and the
limits of the distribution $\gamma_{\rm min}$ and $\gamma_{\rm max}$
as it evolves in the physical space is computed according to
Eq.~(\ref{eq:p(sigma)}). AM ignores synchrotron loses, which yields a
fixed power-law index for the whole distribution of Lorentz factors of
the NTPs. The remaining two parameters needed to specify the
distribution function, $\gamma_{\rm min}$ and $\gamma_{\rm max}$ are
computed from the local values of the hydrodynamic variables. On the
one hand, $\gamma_{\rm min}$ follows from Eq.~(\ref{eq:pmin}) and
$\gamma_{\rm max}$ is obtained from the fact that, $C_\gamma$ is
strictly constant in time if the evolution is adiabatic. Also, in
contrast to SPEV, it is necessary to assume a value of $C_\gamma$
everywhere in the simulated region and not only at the injection
region.


\section{Synchrotron radiation and synthetic radio maps}
\label{sec:synchrotron}

\begin{figure*}
\plotone{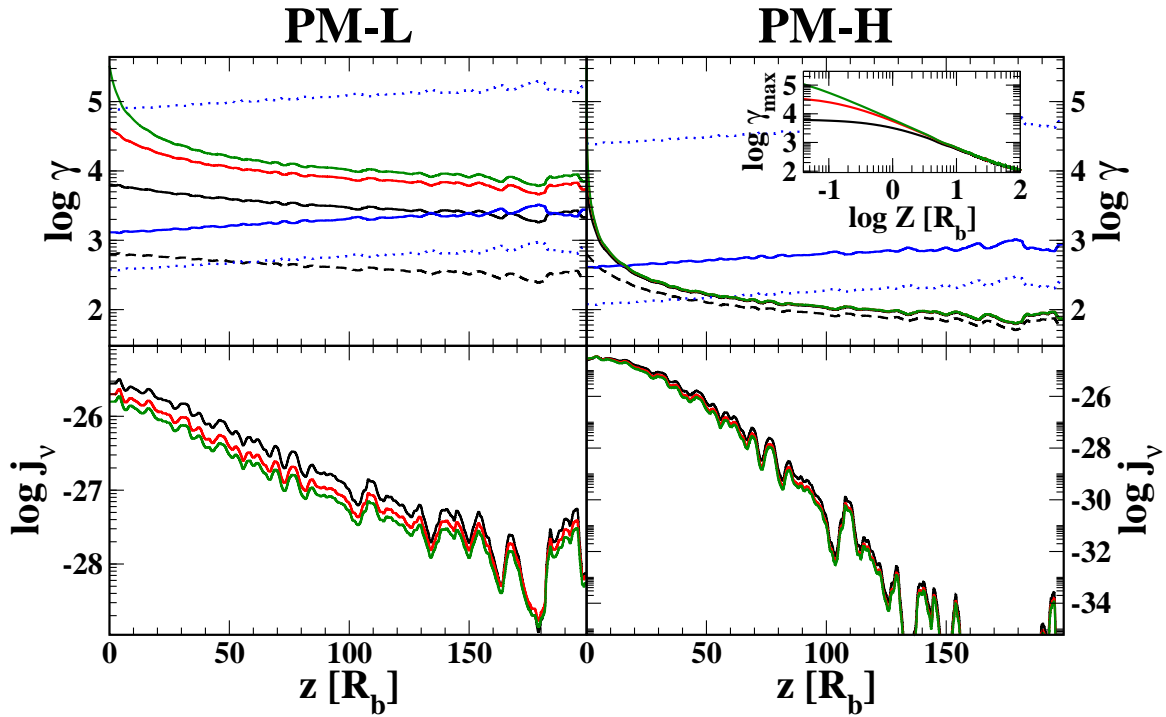}
\caption{Left and right panels correspond to 
  quiescent {\cblue models PM-L and PM-H, respectively}.
  The upper panels display the spectral energy evolution
  along the jet axis.  Thick black, red and green lines show the
  evolution of $\gamma_{\rm max}$ for initial values of $C_\gamma$
  equal to $10$, $10^2$ and $10^3$, respectively. In the upper panels
  we overplot the values of the Lorentz factor corresponding to the
  maximum emission efficiency (solid blue line). The lower and upper
  $\gamma$ where the efficiency drops below 10\% of the maximal as a
  function of distance for an observational frequency of 43~GHz are
  displayed by dotted blue lines above and below the maximum
  efficiency line, respectively. These blue lines have a positive
  slope since the magnetic field decreases with distance, so that ever
  larger Lorentz factors are needed to emit efficiently at a given
  frequency. The lower panels show the synchrotron emissivity of each
  model. The inset in the right panel shows the logarithm of
  $\gamma_{\rm max}$ as a function of the logarithm of $Z$ for the
  three models. It can be seen, that for the model {\cblue PM-H,}
  $\gamma_{\rm max}$ becomes virtually independent of its
  initial value at a distance larger than $\approx 1\,R_b$ from the
  jet nozzle. Please, note the difference in the scales of emissivity
  for the lower panels.
  \label{fig:fiduc_gmax}}
\end{figure*}
\begin{figure*}
\plotone{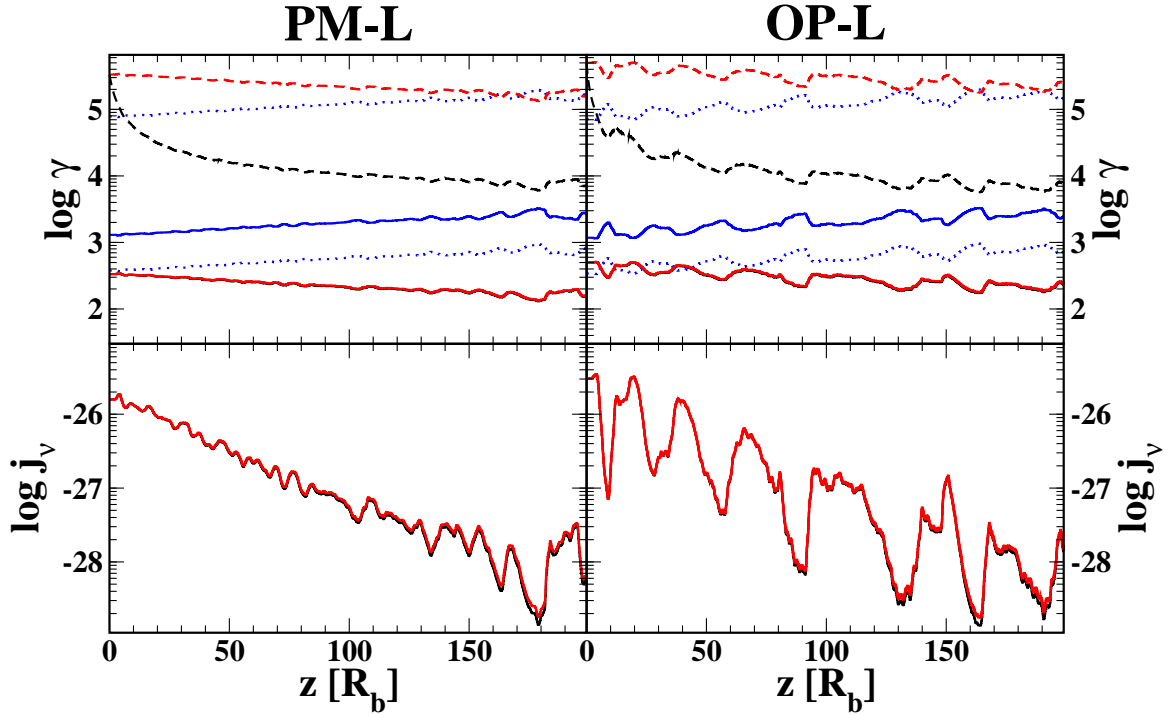}
\caption{In the whole figure, black (red) lines correspond to models
  computed with the SPEV (SPEV-NL) method. The left and right panels
  display properties of the {\cblue PM-L and OP-L} models, respectively. {\it Upper
    panels}: Spectral energy evolution along the jet axis for the
  stationary jet models. The values shown are computed in the jet
  comoving frame but the distance along the axis is measured in the
  laboratory frame (attached to the jet nozzle). The parameters of the
  spectral distribution of NTPs are the same as the ones of the
  reference model (\S~\ref{sec:radio}). In the upper panel the thick
  lines track the values of $\gamma_{\rm min}$ (solid) and
  $\gamma_{\rm max}$ (dashed) of SPEV and SPEV-NL electron
  distributions as a function of the distance along the jet axis. Note
  that the lines corresponding to the values of $\gamma_{\rm min}$ for
  SPEV and SPEV-NL are almost indistinguishable. The curvature in the
  line corresponding to $\gamma_{\rm max}^{\rm SPEV}$, specially in
  the first $30R_b$, shows the effects of synchrotron cooling of the
  highest-energy SPEV particles. { The blue lines have the same
    meaning as in Fig.~\ref{fig:fiduc_gmax}.}  {\it Lower panels}:
  Synchrotron emissivity (Eq.~\ref{eq:j(nu)}) at the jet axis is shown
  as a function of the distance to the jet nozzle. For the parameters
  chosen, most of the electrons of both SPEV and SPEV-NL distributions
  emit synchrotron radiation efficiently in the whole jet. This makes
  that both, SPEV-NL and SPEV methods display a very similar
  emissivity dependence with distance along the jet axis (both curves
  are almost coincident).
  \label{fig:fiduc_distrib}}
\end{figure*}
\begin{figure}
\plotone{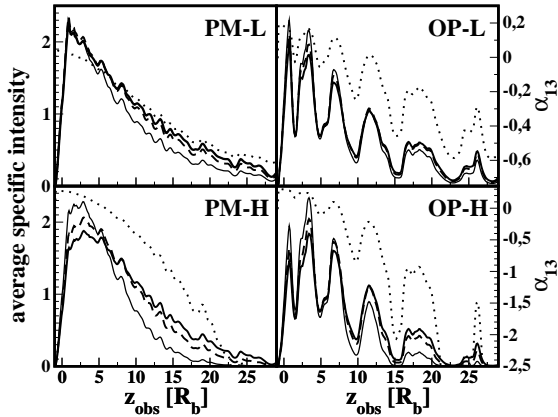}
\caption{{\cblue {\it Upper panels}: Left and right panels display
    properties (as seen by a distant observer) of the PM-L and OP-L
    models, respectively, computed with the SPEV method. The
    thin-solid, dashed and thick solid lines correspond to the
    specific intensity at frequencies 43\,GHz, 22\,GHz and 15\,GHz
    respectively. The intensities are obtained directly from the
    models without convolving the data. For clarity, all the specific
    intensities are normalized to a common value. The dotted line
    shows the spectral index $\alpha_{13}$.  {\it Lower panels}: Same
    as the upper panels, but for the models PM-H and OP-H.
  \label{fig:thick_axis}}}
\end{figure}
\begin{figure}
\plotone{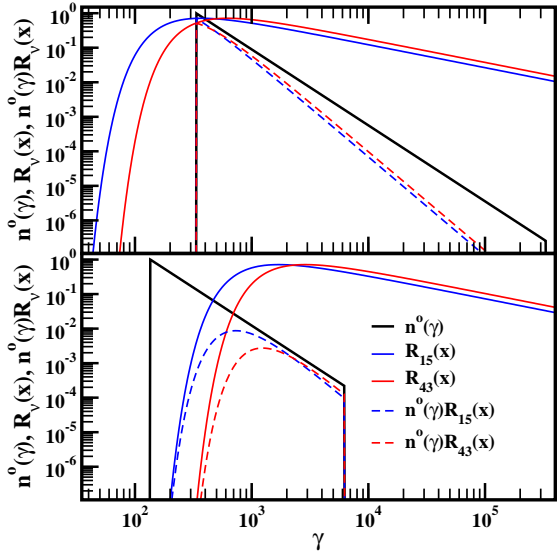}
\caption{The number of particles per unit of energy and unit of volume
  $n^0(\gamma)$ is displayed with a solid black line as a function of
  $\gamma$ {\cblue for the PM-L model}. We omit the temporal dependence
  of $n^0(\tau, \gamma)$ in Eq.~(\ref{eq:n^0_powerlaw}) because we are
  considering quiescent jet models. The solid blue and red lines show
  the synchrotron function $R_\nu(x)$ (Eq.~\ref{eq:R(x)}) at
  frequencies 15\,GHz and 43\,GHz, respectively, while the products
  $n^0(\gamma)R_{15}(x)$ and $n^0(\gamma)R_{43}(x)$ are displayed with
  dashed blue and red lines, respectively. The later products are
  precisely the integrand of the synchrotron emissivity
  (Eq.~\ref{eq:j(nu)}). {\it Lower panel:} Corresponds to the typical
  conditions one encounters downstream the jet {\cblue ($\gamma_{\rm
      min}\sim 135$, $\gamma_{\rm max}\sim 6\times10^3$, $b\sim
    0.002$\,G)}. {\it Upper panel:} Corresponds to the conditions
  found close to the injection nozzle {\cblue ($\gamma_{\rm min}=330$,
    $\gamma_{\rm max}=3.3\times10^5$, $b_b\sim 0.02$\,G)}.
  \label{fig:boundary_effects}}
\end{figure}
\begin{figure}
\plotone{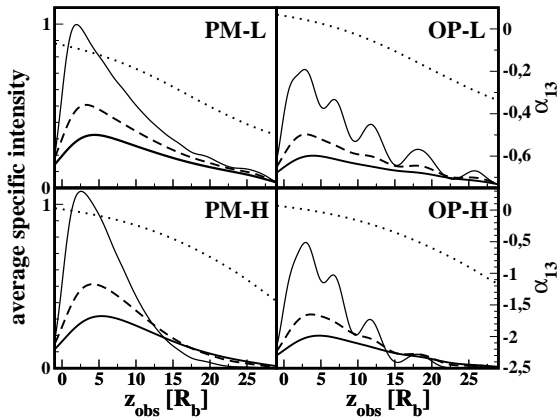}
\caption{Same as Fig.~\ref{fig:thick_axis}, but in this case the
  specific intensities are obtained by convolving the data with a
  circular Gaussian beam whose radius at FWHM are $2.25 R_b$,
  $4.40R_b$ and $6.45R_b$ at frequencies 43\,GHz, 22\,GHz and 15\,GHz,
  respectively. With this convolution we degrade the resolution of our
  data to limits comparable with VLBI
  observations.
\label{fig:thick_axis_conv}}
\end{figure}

The synchrotron emissivity, at a time $\tau$, of an ensemble of NTPs
advected by a thermal plasma element, can be cast in the following
general form (valid both for ordered and random magnetic fields; see
\citealp{Mimica04})
\begin{eqnarray} 
j(\tau, \nu) = \frac{\sqrt{3}e^3 b_{\perp}}{4\pi m_{\rm e} c^2}
\sum_{i=1}^{N_{\rm b}}
  \int_{\gamma_i(\tau)}^{\gamma_{i+1}(\tau)} \der \gamma n^0(\tau, \gamma) g \left( \frac{\nu}{\nu_{\perp}
      \gamma^2} \right) \, ,
\label{eq:j(nu)} 
\end{eqnarray}
where $(g(x), b_\perp,\nu_\perp ) = (R(x), |{\bf b}|, \nu_0)$ if ${\bf
  b}$ is randomly oriented, or $(g(x), b_\perp,\nu_\perp ) = (F(x),
|{\bf b}|\sin\alpha, \nu_0\sin\alpha)$ in case ${\bf b}$ is
ordered. $\alpha$ is the angle the comoving magnetic field forms with
the line of sight, and $\nu_0=3e|{\bf b}|/4\pi m_{\rm e}c$. In the
previous expressions, $F$ is the first synchrotron function
\begin{equation}
 F(x) = x\int_x^\infty d\xi K_{5/3}(\xi), 
\label{eq:F(x)}
\end{equation}
$K_{5/3}$ being the modified Bessel function of index $5/3$, and $R$
is defined as
\begin{equation}
 R(x) :=\frac{1}{2}\int_0^\pi \der \alpha \sin^2\alpha\ \
F\left(\frac{x}{\sin\alpha}\right)\, . 
\label{eq:R(x)}
\end{equation}

The synchrotron self-absorption process is also included in our
algorithm. Thus, we need to compute the synchrotron absorption
coefficient, at a time $\tau$, of an ensemble of NTPs advected by a
thermal plasma element, which can be cast in the following general
form
\begin{eqnarray} 
  \kappa(\tau, \nu) &=& \frac{\sqrt{3}e^3 b_{\perp}}{8\pi m_{\rm e}^2
    c^2 \nu^2} \times \label{eq:kappa(nu)}\\[2mm] 
& &  \sum_{i=1}^{N_{\rm b}}
  \int_{\gamma_i(\tau)}^{\gamma_{i+1}(\tau)} \der \gamma
  \left[-\gamma^2\frac{\der}{\der \gamma} \left(\frac{n^0(\tau,
        \gamma)}{\gamma^2} \right)
  \right] g \left( \frac{\nu}{\nu_{\perp}
      \gamma^2} \right) \, ,\nonumber
\end{eqnarray}

In order to perform the integrals of Eq.~(\ref{eq:j(nu)}) and
(\ref{eq:kappa(nu)}), it is necessary to make some assumption about
the internal distribution of NTPs within each Lorentz factor bin
$i$. As explained in Sect.~\ref{sec:discretization}, we choose to
assume that NTPs distribute as power-law (Eq.~\ref{eq:n^0_powerlaw})
inside of each bin. This choice agrees with the common assumptions
made in the literature and is also supported by theoretical arguments
and observations of discrete radio sources (e.g.,
\citealp{Pacholczyk70}, chapt. 6; \citealp{Koenigl81}), and by
numerical simulations (e.g., \citealp{AGKG01}). Furthermore, it allows
us to build a very efficient and robust method for computing the local
synchrotron emissivity and the local absorption coefficient.  It
consist of tabulating the functions $F(x)$ and $R(x)$, and then
tabulating integrals over power-law distributions of
particles. Proceeding in this way is $\sim 100$ times faster than
computing Eqs.~(\ref{eq:j(nu)})-(\ref{eq:R(x)}) by direct numerical
integration.

The synchrotron coefficients (Eqs.~\ref{eq:j(nu)} and
\ref{eq:kappa(nu)}) of steady jet models result from the time
evolution of the Lagrangian particles injected at the jet nozzle and
spatially transported along the whole jet (the larger the number of
Lagrangian particles, the better coverage of the whole jet). In our
simulations around $N_{\rm steady}=32$ of such Lagrangian particles
(i.e., about 4 particles per numerical cell at the injection nozzle)
are sufficient to properly cover a quiescent jet. If the jet is not
steady, e.g., because a hydrodynamic perturbation is injected, we need
to follow many more Lagrangian particles. It becomes necessary to have
particles everywhere the quiescent jet is perturbed. For the models in
this paper, this means to inject new Lagrangian particles through the
jet nozzle at all time steps after a hydrodynamic perturbation
is set in. The distribution function of the NTPs injected with the
perturbation is the same as that of the particles injected in the
quiescent jet. This is justified since the perturbation only changes
the bulk Lorentz factor, but not the pressure, or the density of the
fluid. In the simulations where we have injected a hydrodynamic
perturbation this implies that we must follow the evolution of more
than $N_{\rm steady} \times N_{\rm timesteps} \gsim 10^5$ Lagrangian
particles. This makes our SPEV simulations effectively
four-dimensional (two spatial, one momentum and a -huge- number of
Lagrangian particles dimension). Therefore, the spatial resolution
that we may afford results severely limited.

The synchrotron coefficients depend on the magnetic field strength and
orientation as well as on the spectral energy distribution
$n^0(\tau,\gamma)$. In our models the magnetic field is {\cblue
  dynamically negligible, thus we set it} up {\it ad hoc}. We choose
that {\cblue $U_{\rm B}$} remains a fixed fraction of the particle energy
density and that the field is randomly oriented.

Synthetic radio maps are build by integrating the transfer equations
for synchrotron radiation along rays parallel to the line, accounting
for the appropriate relativistic effects (time dilation, Doppler
boosting, etc.). The technical details relevant for this procedure can
be found in Appendix~\ref{appendix:imaging}.

\section{Radio emission}
\label{sec:radio}

The goals of this section are twofold. First, we validate the new
algorithm comparing the synthetic radio maps obtained with SPEV
without accounting for synchrotron losses with the {\cblue ones
  obtained} using AM. For this purpose, we will employ the SPEV method
to evolve NTPs but taking $k_{\rm s}=0$ in
Eq.~(\ref{eq:dpdsigma2}). We will refer to this method of evaluating
the evolution of NTPs as SPEV-NL. Second, we will show the differences
induced by accounting for synchrotron losses in the evolution of NTPs.

\subsection{Calibration of the method}
\label{sec:calibration1}

In order to properly compare SPEV-NL and AM results we set up the same
spectral parameters at the jet nozzle for both: $q_1= 2.2$,
$\gamma_{\rm min}=330$, $C_\gamma=10^3$ and $\rho_a =
2\times10^{-21}\,$g\,cm$^{-3}$. We produce all our images for a
canonical viewing angle of $10^\circ$ and assuming that $R_b=0.1\,$pc.
The comoving magnetic field strength is ${\rm b}_b:=\sqrt{{\bf
    b}^2}=0.02\,$G {\cblue (model PM-L-NL) and 0.03\,G (model OP-L-NL)}.%
\footnote{\cblue With such values of ${\rm b}_b$ the magnetic field is
  dynamically negligible.}

For the set of reference parameters we have considered, the synthetic
radio maps of the quiescent jets produced with SPEV-NL yield very
small differences with respect those computed with AM
(Fig.~\ref{fig:fiduc-emiss}). Indeed, the overall agreement between
both methods in the predicted quiescent radio maps is remarkably good,
particularly, if we consider the fact that SPEV is a Lagrangian method
while AM is Eulerian.

Looking at the synthetic radio maps of model {\cblue OP-L-NL}
(Fig.~\ref{fig:fiduc-emiss}), we observe, as in G97, a regular pattern
of knots of high emission, associated with the increased specific
internal energy and rest-mass density of internal oblique shocks
produced by the initial overpressure in this model. The intensity of
the knots decreases along the jet due to the expansion resulting from
the gradient in external pressure. Some authors (e.g., \citealt{DM88};
G95; G97; \citealt{Marscheretal08}) propose that the VLBI cores may
actually correspond to the first of such recollimation shocks.  Since,
for the parameters we use, the source absorption for frequencies above
1\,GHz is negligible, the jet core reflects the ad hoc jet inlet in
the {\cblue PM-L-NL} model, while we shall associate it with the first
recollimation shock for model {\cblue OP-L-NL}. The rest of the knots are
standing features in the radio maps for which, there exists robust
observational confirmation \citep{Gomez02,Gomez05}.

Since the synchrotron losses affect more the higher energy part of the
distribution of NTPs than the lower one, we have also validated our
code by considering the dependence of the results with the limit
$\gamma_{\rm max}$ and checked them against the theoretical
expectations (e.g., \citealp{Pacholczyk70}).  For this we reduce the
value of $\gamma_{\rm max}$ keeping all other parameters fixed and
equal to those of the {\cblue PM-L} model. Since the value of
$\gamma_{\rm max}$ is set by the ratio $C_\gamma$, in order to study
the dependence of the results with $\gamma_{\rm max}$, we have
computed a set of models combining three different values of
$C_\gamma=\{10^3,10^2,10\}$ and ${\rm b}_b=0.02$\,G. Additionally, to
highlight the effect of the radiative losses, we have performed the
same simulations (varying $C_\gamma$) for a larger value of the beam
magnetic field, {\cblue equal to that of the model PM-H}.

\begin{figure}
\plotone{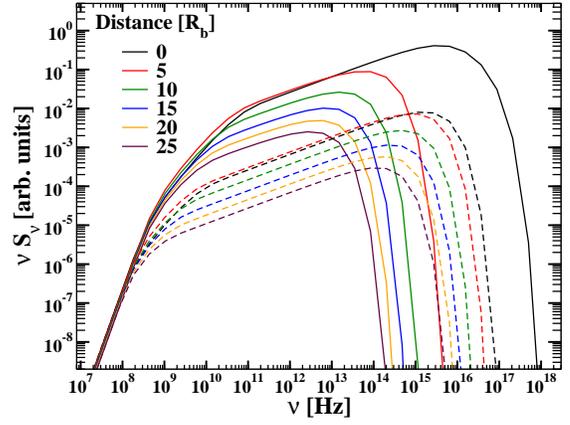}
\caption{The different lines in the plot show the spectral energy
  distribution of selected points along the jet axis of model PM with
  two different magnetic fields. All the models have been computed
  using the SPEV method with the reference parameters of
  \S~\ref{sec:radio}. Solid lines correspond to the {\cblue PM-L model}
  and dashed lines to the {\cblue PM-S} model. The distance to the nozzle (in $R_b$) to
  which each spectrum corresponds is provided in the
  legend. Synchrotron self-absorption is dominant at frequencies below
  few hundred MHz.
  \label{fig:spec-axis}}
\end{figure}

\begin{figure*}
\plotone{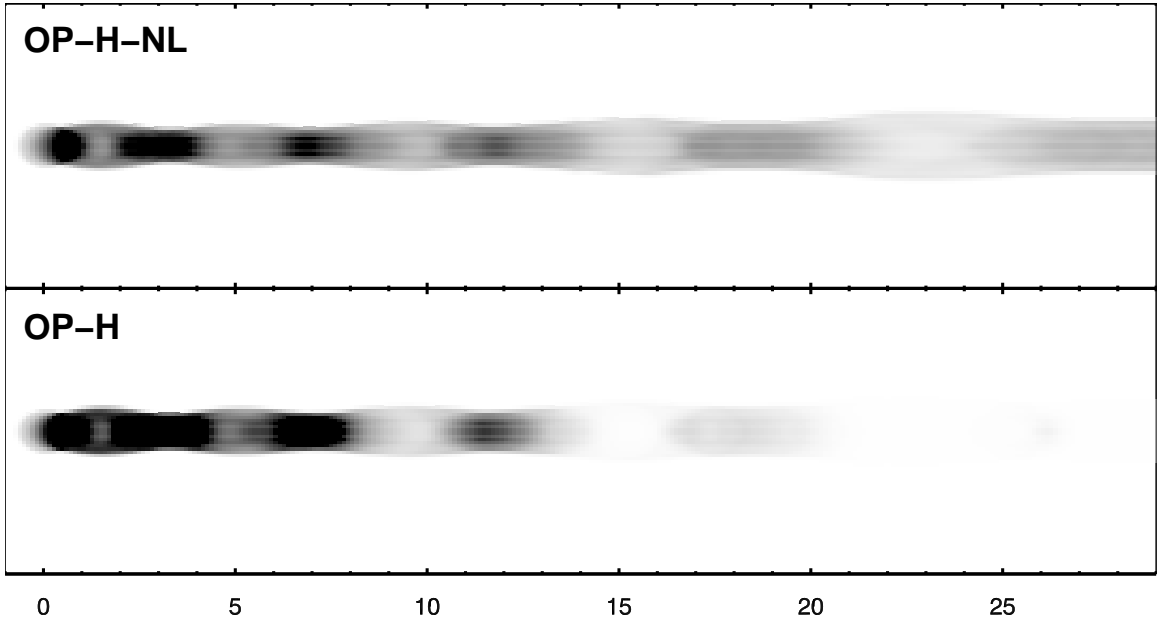}
\caption{Same as Fig.~\ref{fig:fiduc-emiss}, but in this case {\cblue for the OP-H model}.
  \label{fig:highB-emiss}}
\end{figure*}
\begin{figure*}[htb]
\plotone{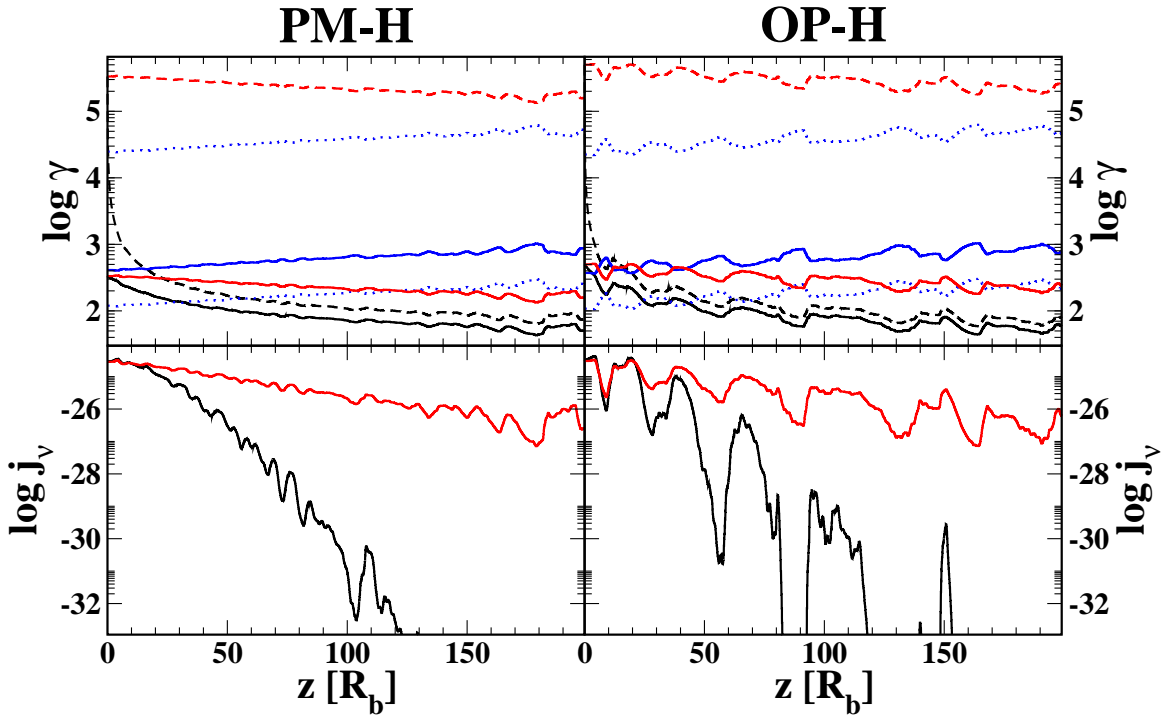}
\caption{Same as Fig.~\ref{fig:fiduc_distrib} but for {\cblue the PM-H and OP-H models}. In this case, the synchrotron
  losses in SPEV are so important that, $\gamma_{\rm max}^{\rm SPEV}$
  leaves the efficient synchrotron radiation regime. The point where
  this happens depends on the model. The crossing of the $\gamma_{\rm
    max}^{\rm SPEV}$ with the line denoting a 10\% synchrotron
  efficiency limit occurs when the particles reach $z \sim 50R_b$ for
  model PM-H and much earlier $z \sim 24R_b$ in case of model OP-H. After
  this line crossings the whole distribution radiates very little at
  the considered observational frequency.  This produces a substantial
  dimming of SPEV models at large $z$. The difference in the
  emissivity as a function of the distance $z$ is larger for the PM-H
  model than for the OP-H model because of the re-compressions that the
  SPEV-NTP experiences at shocks. 
  \label{fig:high-B}}
\end{figure*}

For {\cblue model PM-L} (Fig.~\ref{fig:fiduc_gmax}, left panel),
radiative losses are negligible, and the reduction in $C_\gamma$
(i.e., in $\gamma_{\rm max}$), does not change appreciably the radio
maps at radio observing frequencies. Indeed, except the model with the
lowest $C_\gamma$ (corresponding to $\gamma_{\rm max}=2200$) beyond
$160\,R_b$, all the models stay above the 100\% efficient radiation
limit along the whole jet.

The models with larger magnetic field ${\rm b}_b=0.2\,$G
(Fig.~\ref{fig:fiduc_gmax}, right panel), undergo a much faster
evolution. The emissivity along the jet axis drops very quickly and at
$z=150R_b$, {\cblue it} is five orders of magnitude smaller than
{\cblue for the PM-L model}. After a very short distance ($\simeq
1\,R_b$), synchrotron losses bring $\gamma_{\rm max}$ of all three
models to a common value which is independent of the initial one (note
that the variation of $\gamma_{\rm max}$ with distance is
indistinguishable for the three models except in the zoom displayed in
the inset of the top right panel of Fig.~\ref{fig:fiduc_gmax}). The
reason for this degenerate evolution resides in the relatively large
magnetic field strength {\cblue (see \citealp{Pacholczyk70},
  Eq.~6.20)}.
%
%
Thus, our method is able to reproduce the common evolution
of models with different values of $\gamma_{\rm max}$ and a relatively
large magnetic field.

\subsection{On the relevance of synchrotron losses}
\label{sec:synlos}

Having verified that our method (SPEV-NL) compares adequately to the
AM, we now turn to the specific role that synchrotron losses play in
the evolution of NTPs. For that, we compare in
Fig.~\ref{fig:fiduc_distrib} the spectral properties of NTPs in
quiescent jet models using both SPEV-NL and SPEV methods. It is
obvious that the highest energy particles of the distribution cool
down rather quickly (see the fast decay of the dashed black curves in
the upper panels of Fig.~\ref{fig:fiduc_distrib}) even for the small
value of ${\rm b}_b$ considered here. Most of the spectral evolution
triggered by synchrotron cooling at high values of $\gamma$ happens in
the first $25R_b-50R_b$. After that location, the ratio $C_\gamma$ is
much smaller than at the injection nozzle ($C_\gamma\lsim 50$), and
the evolution of the NTP population is dominated by the adiabatic
cooling/compression downstream the jet. In contrast, the upper limit
of the SPEV-NL distribution (dashed red curves in the upper panels of
Fig.~\ref{fig:fiduc_distrib}) only changes by a factor of 2 along the
whole jet length. Theoretically, it is well understood that it is
possible to undergo a substantial spectral evolution (triggered by
synchrotron losses) and, simultaneously, not to have any manifestation
of such evolution at radio frequencies
\citep[e.g.,][]{Pacholczyk70}. The substantial decrease of
$\gamma_{\rm max}$ triggered by the radiative losses, does not affect
much the value of the integral that has to be performed over $\gamma$
in order to compute the emissivity in Eq.~(\ref{eq:j(nu)}), since most
of the emitted power at radio-frequencies happens relatively close to
$\gamma_{\rm min}$, where synchrotron losses are
negligible. Certainly, at higher observing frequencies this is not the
case, and the emissivity substantially drops because of the fact that
both, the synchrotron losses (Eq.~\ref{eq:dpdt_syn}) and the frequency
at which the spectral maximum emission is reached depend on the square
of the non-thermal electron energy (and on the magnetic field
strength).

  We define the spectral index between two radio frequencies as
\begin{eqnarray} \label{eq:alpha_ij}
  \alpha_{ij} = \frac{\log{(S_i/S_j)}}{\log{(\nu_i/\nu_j)}},
\end{eqnarray}
where $S_i$ and $S_j$ are the flux densities at the frequencies
$\nu_i$ and $\nu_j$, respectively. Since we compute synthetic radio
maps at three different radio frequencies ($\nu_1=15\,$GHz,
$\nu_2=22\,$GHz, and $\nu_3=43\,$GHz), we may define three different
spectral indices. For convenience, in the following, we consider the
spectral index $\alpha_{13}$ between 15\,GHz and 43\,GHz. Furthermore,
we may compute $\alpha_{13}$ for both {\it convolved} or {\it
  unconvolved} flux densities. The unconvolved flux density is
directly obtained from the simulations and has an extremely good
spatial resolution, viz. the unconvolved radio images have a
resolution comparable to that of the hydrodynamic data. The convolved
flux densities result from the convolution with a circular Gaussian
beam of the unconvolved data. The FWHM of the Gaussian beam is
proportional to the observing wavelength. This convolution is
necessary to degrade the resolution of our models down to limits
comparable with typical VLBI observing resolution. We note that in
order to compute the spectral index for convolved flux densities, we
have to employ the same FWHM convolution kernel for the data at the
two frequencies under consideration. Thus, to compute $\alpha_{13}$
for convolved data, we employ the same Gaussian beam with a FWHM
6.45\,$R_b$ for both flux densities at 15\,GHz and 43\,GHz.

Our models are computed for an electron spectral index $q=q_1=2.2$. We
verify that, at large distances to the jet nozzle, unconvolved models
(Fig.~\ref{fig:thick_axis} upper panels) tend to reach the expected
value $\alpha=(1-q)/2=-0.6$ for an optically thin source. This
asymptotic value is reached smoothly in case of the {\cblue PM-L and
  PM-H models} and it is modulated by the presence of inhomogeneities
(recollimation shocks) in the beam of models {\cblue OP-L and OP-H}.

%
%
Close to the jet nozzle, our unconvolved models display flat or even
inverted ($\alpha_{13}>0$) spectra (Fig.~\ref{fig:thick_axis}), in
spite of the fact that the jets are optically thin throughout their
whole volume. The occurrence of flat or inverted spectrum depends on
the magnetic field strength and differs for OP and PM models.  {\cblue As
  shown in Fig.~\ref{fig:thick_axis}, the PM-L} model shows an
inverted spectrum for $z\lsim2.5R_b$, {\cblue while} the {\cblue OP-L} model
displays a pattern of alternated inverted and normal ($\alpha_{13}<0$)
spectra for $z \lsim 12.5R_b$. The spectral inversion in the {\cblue
  OP-L} model happens where standing features (associated to
recollimation shocks in the beam) are seen in the jet.

If synchrotron losses are not included, the spectral behavior of
models {\cblue PM-L and OP-L} remains unchanged, because in such a case
loses are negligible. However, if for the models {\cblue PM-H and OP-H}
the losses are not accounted for (which is, obviously, a wrong
assumption), the jet displays an inverted spectrum up to distances $z
\sim 30R_b$.

The behavior of the spectral index exhibited by our models close to
the jet nozzle contrasts with the theoretical expectations for an
inhomogeneous, optically thin jet with a negative electron spectral
index, for which the jet inhomogeneity is predicted to steepen the
spectrum \citep[e.g.,][]{Marscher80,Koenigl81}. To explain this
discrepancy we argue that the analytic predictions are based on the
assumption that the limits of the energy distribution of the NTPs
safely yield that the contribution of the synchrotron functions
(Eqs.~\ref{eq:F(x)} and \ref{eq:R(x)}) to the synchrotron coefficients
(Eqs.~\ref{eq:j(nu)} and \ref{eq:kappa(nu)}) is proportional to some
power of the frequency and of the NTP's Lorentz factor. This situation
does not happen if the lower limit of the distribution $n^0(\gamma)$,
$\gamma_{\rm min}$, is (roughly) larger than the value $\gamma_{_{\rm
    M}}$ at which the synchrotron function $R(x_{\rm low})$
(Eq.~\ref{eq:R(x)}) reaches its maximum, where $x_{\rm low}=\nu_{\rm
  low} / \nu_0 \gamma^2$, and $\nu_{\rm low}$ is the smallest
observing frequency in the comoving frame. Since the function $R(x)$
has a maximum for $x\simeq 0.28$, one finds that the condition to have
an inverted spectrum is $\gamma_{\rm min} \gsim \gamma_{_{\rm
    M}}\simeq 1.9 \times (\nu_{\rm low}/ \nu_0)^{1/2} {\cal
  D}^{-1/2}$, where ${\cal D}:=1/\Gamma (1 - v\cos{\theta})$ is the
Doppler factor. Since, in our case, $\nu_{\rm low}=15$\,GHz, we may
also write
\begin{equation}
  \gamma_{\rm min} \gsim 113 \left( \frac{\nu_{\rm low}}{\rm 15\,GHz}
    \right)^{1/2} \left(\frac{b}{\rm 1\,G}\right)^{-1/2} {\cal
      D}^{-1/2}.
\label{eq:gamma_min2}
\end{equation}
\noindent
Figure~\ref{fig:boundary_effects} shows how this boundary effect
substantially modifies the emissivity at 15\,GHz and 43\,GHz for the
model {\cblue PM-L}. At the injection nozzle
(Fig.~\ref{fig:boundary_effects} upper panel) the lower limit of the
integral in Eq.~(\ref{eq:j(nu)}) is set by $\gamma_{\rm min}$ and not
by the lower limit of $R_\nu(x)$. However, downstream the jet
(Fig.~\ref{fig:boundary_effects} lower panel) the situation reverses
and the fast decay of $R_\nu(x)$ for $\gamma<300$ sets the lower limit
of the emissivity integral. {\cblue Thus,} close to the nozzle, the value
of the area below the $n^0(\gamma)R_{43}(x)$ curve, which is
proportional to the emissivity at 43\,GHz, is larger than that below
the curve $n^0(\gamma)R_{15}(x)$. Hence, there is an emissivity excess
at 43\,GHz compared to that at 15\,GHz. As a result, the $\alpha_{13}$
becomes positive close to the jet nozzle. Far away from the nozzle the
emissivity at 15\,GHz almost doubles that at 43\,GHz, explaining why
values of $\alpha_{13}< 0$ are reached {\cblue asymptotically}. 

The convolved models display some traces of the behavior shown for the
uncovolved ones. For example, OP models display a flat or inverted
spectrum very close to the jet nozzle (Fig.~\ref{fig:thick_axis_conv}
right panels). This is not the case for {\cblue PM-L model}
(Fig.~\ref{fig:thick_axis_conv}). Since the resolution of the
convolved data is much poorer than that of the unconvolved one,
$\alpha_{13}$ exhibits a quasi monotonically decreasing profile from
the jet nozzle (where $-0.1\lsim \alpha_{13}\lsim 0$). The coarse
resolution of the convolved data also blurs any signature in the
spectral index associated to the existence of cross shocks in the beam
of OP models. Furthermore, the decay with distance of the spectral
index is shallower for the convolved flux data than for the
unconvolved one. {\cblue Hence,} the theoretical value
$\alpha_{13}=-0.6$, which is expected for an optically thin
synchrotron source, is reached nowhere in the jet models {\cblue PM-L and
  OP-L} (Fig.~\ref{fig:thick_axis_conv}).

As expected, at frequencies below a few hundred MHz, the jet is
strongly self-absorbed everywhere (Fig.~\ref{fig:spec-axis}). Close to
the jet nozzle, there is not a clear turnover frequency between the
self-absorbed part of the spectrum and the optically thin
one. Instead, we observe a smooth transition between both regimes. Far
from the nozzle, the self-absorption turnover is much more peaked. It
is known \citep{TK07} that in contrast with a distribution of NTP that
follows a power-law extending to $\gamma_{\rm min}\gsim 1$, if the
power-law is restricted to a relatively large, but not unrealistic
$\gamma_{\rm min}$, or if the electron distribution {\cblue was}
monoenergetic, the intensity can be flat over nearly two decades in
frequency (which implies that the energy flux grows linearly over the
same frequency range). Our {\cblue PM-L} models have $\gamma_{\rm
  min}=330$ at the injection nozzle and reduce it to $\gamma_{\rm
  min}\simeq 200$ at $z=200R_b$ because of the adiabatic expansion of
the jet (Fig.~\ref{fig:fiduc_distrib} upper left). As we have argued
in \S~\ref{sec:radio} close to the jet nozzle, $\gamma_{\rm
  min}\gsim\gamma_{_{\rm M}}$, which means that $\gamma_{\rm min}$ is
sufficiently large to be in the range where a smooth turnover
transition is expected, in agreement with \cite{TK07}. Far away from
the nozzle, since $\gamma_{\rm min}$ decreases, we recover the more
standard situation in which an obvious turnover frequency can be
identified.

{\cblue Provided} that close to the nozzle our PM (also OP) models are
weakly self-absorbed ({\cblue at 15\,GHz, the solid black line in
  Fig.~\ref{fig:spec-axis} has not reached the power-law regime yet}),
one may question whether the spectral inversion we have found is not
also the result of opacity effects. We have dismissed such a
possibility by running models with the SPEV method {\cblue including
  no} absorption.

\subsubsection{Dependence with the magnetic field strength}
\label{sec:magfield}

In order to {\cblue study the effect of} intense synchrotron losses we
{\cblue consider models PM-H and OP-H} (Figs.~\ref{fig:highB-emiss} and
\ref{fig:high-B}). Very close to the injection nozzle ($Z\sim
50\,R_b$) the line denoting the evolution of $\gamma_{\rm max}^{\rm
  SPEV}$ crosses the line corresponding to a 10\% synchrotron
efficiency limit (lower blue thick line; Fig.~\ref{fig:high-B}) and
most of the synchrotron emissivity falls outside of the observational
frequency. {\cblue Because of a} stronger magnetic field {\cblue than in
  models PM-L and OP-L}, more energy is lost close to the jet nozzle
than far from it and, thus, SPEV-radio-maps look much shorter than
SPEV-NL-radio-maps (Fig.~\ref{fig:highB-emiss}). An alternative way to
see such an effect is through the rapid decay of $\gamma_{\rm
  max}^{\rm SPEV}$ in the first $10R_b$ in Fig.~\ref{fig:high-B},
right panel. Afterwards, the adiabatic changes dominate the NTP
evolution. The initial period of fast evolution is even shorter if a
larger magnetic field {\cblue were to be} considered.

The intensity contrast between shocked and unshocked jet regions of
{\cblue model OP-H (Fig.~\ref{fig:high-B}) is larger than that of
  model OP-H-NL}. Indeed, the {\cblue OP-H} model appears as a
discontinuous jet (Fig.~\ref{fig:highB-emiss}) because of the slightly
larger intensity increase than in the {\cblue OP-H-NL} model when the
NTP distribution passes through cross-shocks and the much more
pronounced intensity decrease at rarefactions. We note that, although
the adiabatic evolution is followed with the same algorithm in SPEV
and SPEV-NL, the radiative losses change substantially the NTP
distribution that it is injected through the nozzle after very short
distances. The consequence being that the NTP distribution
$n^0(\tau,\gamma)$ that faces shocks and rarefactions downstream the
nozzle is rather different when using SPEV or SPEV-NL method and,
therefore, the relative intensity of shocked and unshocked regions is
also different depending on whether synchrotron losses are included or
not in the calculation.

The outlined differences between {\cblue models OP-H and OP-H-NL (with
  shocks in the beam)}, have to be interpreted with caution since none
of the {\cblue methods accounts} for the injection of high-energy
particles at shocks. But independent of this, we expect that if the
magnetic field is sufficiently large, the SPEV method will yield a
rather fast evolution of such particles and, thereby, a faster decay
of the intensity downstream the shock.

The most relevant difference between the upper and lower panels of
Fig.~\ref{fig:highB-emiss}, is the increased brightness of the jet
close to the injection nozzle and the steeper fading of the jet when
energy losses are included. This fact poses the paradox that the
method that accounts for radiative losses (SPEV) yields brighter
standing features close to the injection nozzle (far from the nozzle
the situation reverses and the SPEV-NL model is brighter than SPEV
one). In order to disentangle this apparent contradiction, we shall
consider that the plasma is compressed at standing shocks, which
yields a growth of the magnetic field energy density (proportional to
the pressure in our case), and triggers a faster cooling of the
high-energy particles. Since the SPEV method conserves the number
density of NTPs (Eq.~\ref{eq:N}), due to the synchrotron losses,
high-energy particles reduce their energy and accumulate into an
interval of Lorentz factor which is smaller than in the case of
SPEV-NL models.  As in such reduced Lorentz factor interval NTPs
radiate more efficiently at the considered radio-frequencies, the
emissivity of SPEV models at strong compressions (like, e.g., the
considered cross shocks) becomes larger than that corresponding to
models which do not include synchrotron losses. It is important to
note that this situation happens in our models rather close to the jet
nozzle. The reason being that after the NTPs have suffered a
substantial synchrotron cooling, the evolution of the NTP distribution
is dominated by the adiabatic terms of Eq.~\ref{eq:dpdsigma2}. In such
a regime, reached by our models at a certain distance from the jet
nozzle, the evolution of SPEV-NL and SPEV models is qualitatively
similar. Considering the different qualitative evolution of the NTP
distribution close to the nozzle and far from it, we refer to such
epochs as {\it losses-dominated} and {\it adiabatic} regimes,
respectively. These terms agree with the commonly used in the
literature to refer to similar phenomenologies (e.g., \citealt{MG85}).

For {\cblue PM-H and OP-H models}, the spectral behavior is dominated
by the change of slope of the NTP Lorentz factor distribution beyond
the synchrotron cooling break at $\gamma=\gamma_{\rm
  br}$. Theoretically, an optically thin inhomogeneous jet shall
display a spectral index $\alpha=(q+1)/2$, if the radiation in the
observational band is dominated by the electrons with Lorentz factors
$\gamma\gsim\gamma_{\rm br}$, or $\alpha\simeq-2.7$ if the emission is
dominated by electrons with Lorentz factors close to $\gamma_{\rm
  max}$ \citep{Koenigl81}\footnote{{\cblue çWe obtain this value from}
  the expression $\alpha_{s3}=(m+2-n)/m$ of \cite{Koenigl81} with
  $m=1.15$ and $n=0$. The values of $m$ and $n$ are computed from the
  decay with the distance to the jet nozzle of the magnetic field
  strength { $|{\bf b}|\propto z^{-m}$} and of the number density of
  NTPs per unit of energy { $n^0\propto z^{-n}$},
  respectively.}. Figure~\ref{fig:thick_axis} (lower panels) shows
that asymptotically (viz., at large $z$) unconvolved models reach
values $\alpha_{13}\lsim -2.5$, implying that the highest energy
electrons with $\gamma\sim\gamma_{\rm max}$ are the most efficiently
radiating at the considered observing frequencies. The value of
$\gamma_{\rm max}$ differs significantly when synchrotron loses are
not included. This fact explains the inversion of the spectrum along
the whole jet if synchrotron losses were not included {\cblue (PM-H-NL
  and OP-H-NL models)}. Thereby, synchrotron losses tend to produce a
``normal'' spectrum ($\alpha_{ij}<0$) if the magnetic field is large.

Regarding the convolved data, we note that models with a higher
magnetic field display the same qualitative phenomenology discussed in
\S~\ref{sec:radio}. In this case, the theoretical value
$\alpha\simeq-2.7$ is not reached neither by the {\cblue PM-H}
($\alpha_{13,min}^{\rm PM-H}=-1.4$) nor by the {\cblue OP-H}
($\alpha_{13,min}^{\rm OP-H}=-1.1$) model
(Fig.~\ref{fig:thick_axis_conv} lower panels).

\section{Infrared to X-rays emission}
\label{sec:infrared_to_X}

We have computed the spectral properties of some of our quiescent jet
models above radio frequencies. {\cblue We note, that we have not
  included any particle acceleration process at shocks in the SPEV
  method, thus,} the spectrum beyond infrared frequencies has to be
taken carefully. If any shock acceleration mechanism were included, a
larger contribution of the shocked regions will be present. In
addition, the inverse Compton process, may shape the emission at such
high energies, and such cooling process is presently not included in
SPEV.


{\cblue The results for models PM-S and PM-L (Tab.~\ref{tab:models}),
  which have no or extremely weak shocks are displayed in
  Fig.~\ref{fig:spec-axis}, where we show} the spectral energy
distribution at selected distances from the nozzle for points located
along the jet axis. {\cblue The small magnetic field of model PM-S
  (Fig.~\ref{fig:spec-axis} dashed lines) minimizes the energy losses,
  but also the observed flux in the optical or X-ray band, rendering
  observable at such wavelengths the hydrodynamic jet models
  considered here (if the jet is sufficiently close). In the case of
  model PM-L,} right at the nozzle ($z=0$ in
Fig.~\ref{fig:spec-axis}), the energy flux cut-off is located at
$\simeq 10^{18}\,$Hz. This means {\cblue that, we} could observe the
jet core in the soft X-ray band, if the source was sufficiently
close. However, the core size at such frequencies is very small
{\cblue (as it is expected; see} e.g., \citealt{MG85}). This is
reproduced in our models since at such a short distance as $z=5R_b$,
the jet can barely be observed in the Near Ultraviolet or, perhaps in
the optical band (Fig.~\ref{fig:spec-axis}, red solid line), but there
is virtually no flux in the X-ray band because of the fast NTP cooling
for the considered magnetic field energy density at the jet nozzle. In
the Near Infrared range, the jet could perhaps be observable up to
distances of $10R_b-15R_b$. {\cblue A} larger magnetic field drives
{\cblue a} faster cooling, rendering undetectable the jet even at
infrared frequencies. This phenomenology has been invoked to explain
the relative paucity of optical jets with respect to radio
jets. However, there are a number of authors which claim that a large
proportion of jets generate significant levels of both optical and,
even, X-ray emission (e.g., \citealp{Perlmanetal06}). Our results
shall not be taken in support of any of the two thesis since energy
losses depend also on the magnetic field strength
(Eq.~\ref{eq:dpdt_syn}), which we fix {\it ad hoc}.

\section{Evolution of a superluminal component}
\label{sec:spec_evol_component}

In this section we discuss the time dependent observed emission once a
hydrodynamic perturbation is injected into the jet (see
Sec.~\ref{sec:perturb}). Following the convention of G97, {\cblue we
  call {\it components}} to local increases of the specific intensity
in a radio map, while we use {\it perturbation} to denominate the
variation of the hydrodynamic conditions injected through the jet
nozzle.
%
%
In order to magnify the effect of synchrotron losses in our models, we
{\cblue discuss models PM-H and OP-H} in Sec.~\ref{sec:losses}, and we
also look for the differences between the PM and OP models.%
%
\footnote{{\cblue In the online material we provide a movie
    ("PMOP-fiduc.mpg") where the evolution of the total intensity at
    43\,GHz is displayed for models PM-L and OP-L.}}
While the standing shocks of the beam of model OP-H are very weak, the
shocks developed by the hydrodynamic perturbation are rather
strong. Since we have not included in our method the acceleration of
NTPs at relativistic shocks, computing the synchrotron emission at
frequencies above radio may yield inconsistent results. Therefore, we
only analyze the spectral properties of the emission in radio
bands. Finally we show spacetime plots of hydrodynamic and emission
properties along the jet axis in Sec.~\ref{sec:spacetime_analysis}.

\begin{figure*}
\plotone{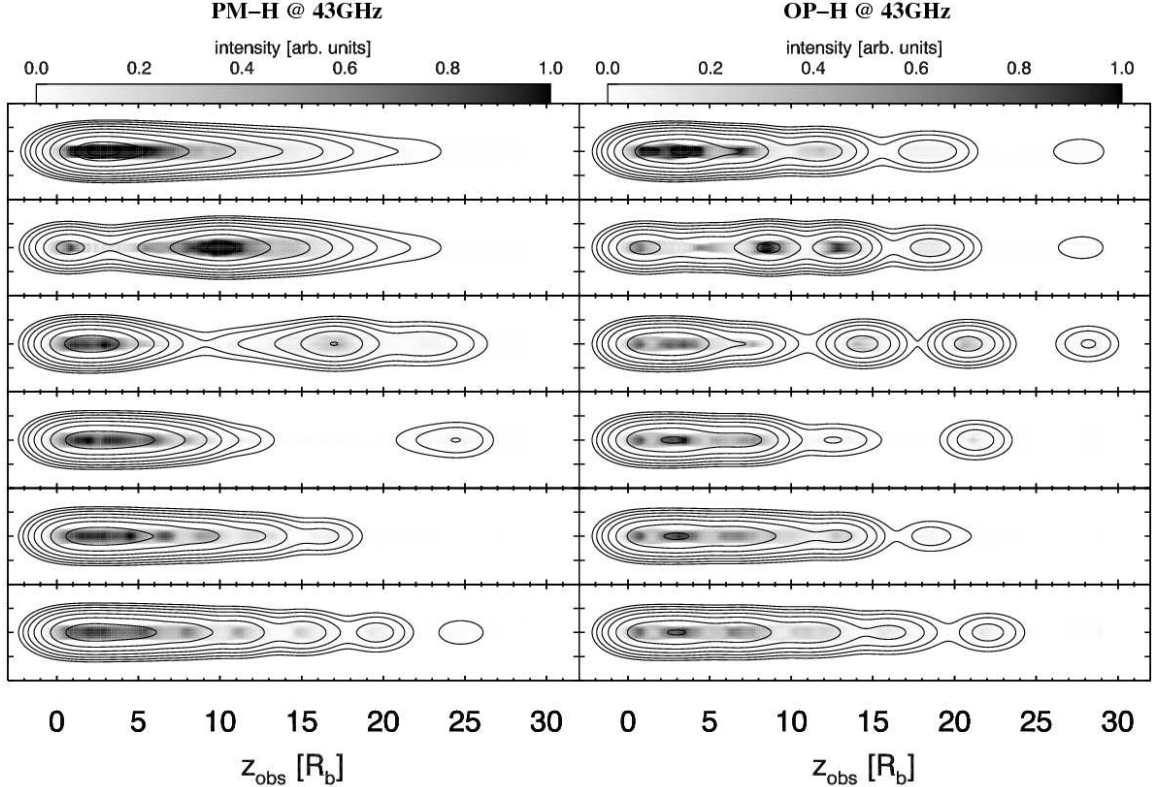}
\caption{{\cblue Snapshots of the emission at 43\,GHz due to component
    evolution computed with the SPEV method, for the PM-H (left) and
    OP-H (right) models. From top to bottom panels show the observed
    emission $0.02$, $0.39$, $0.75$, $1.12$, $1.94$ and $4.58$ years
    after the component appears. The same gray scale has been used for
    all snapshots. The superimposed contours have been obtained by
    convolving the image with a circular Gaussian beam whose radius at
    FWHM is $2.25 R_b$. The contour levels are 0.005, 0.01, 0.02,
    0.04, 0.08, 0.16, 0.32, 0.64 and 0.9 of the maximum of the
    convolved emission. The horizontal length scale is expressed in
    units of $R_b = 0.1\,$pc, while the vertical length scale has been
    compressed and spans only $10R_b$. The main component has moved
    out of the right boundary in the lower two panels.  }
  \label{fig:PMOP-highB-snap}}
\end{figure*}
%


\begin{figure*}
\plotone{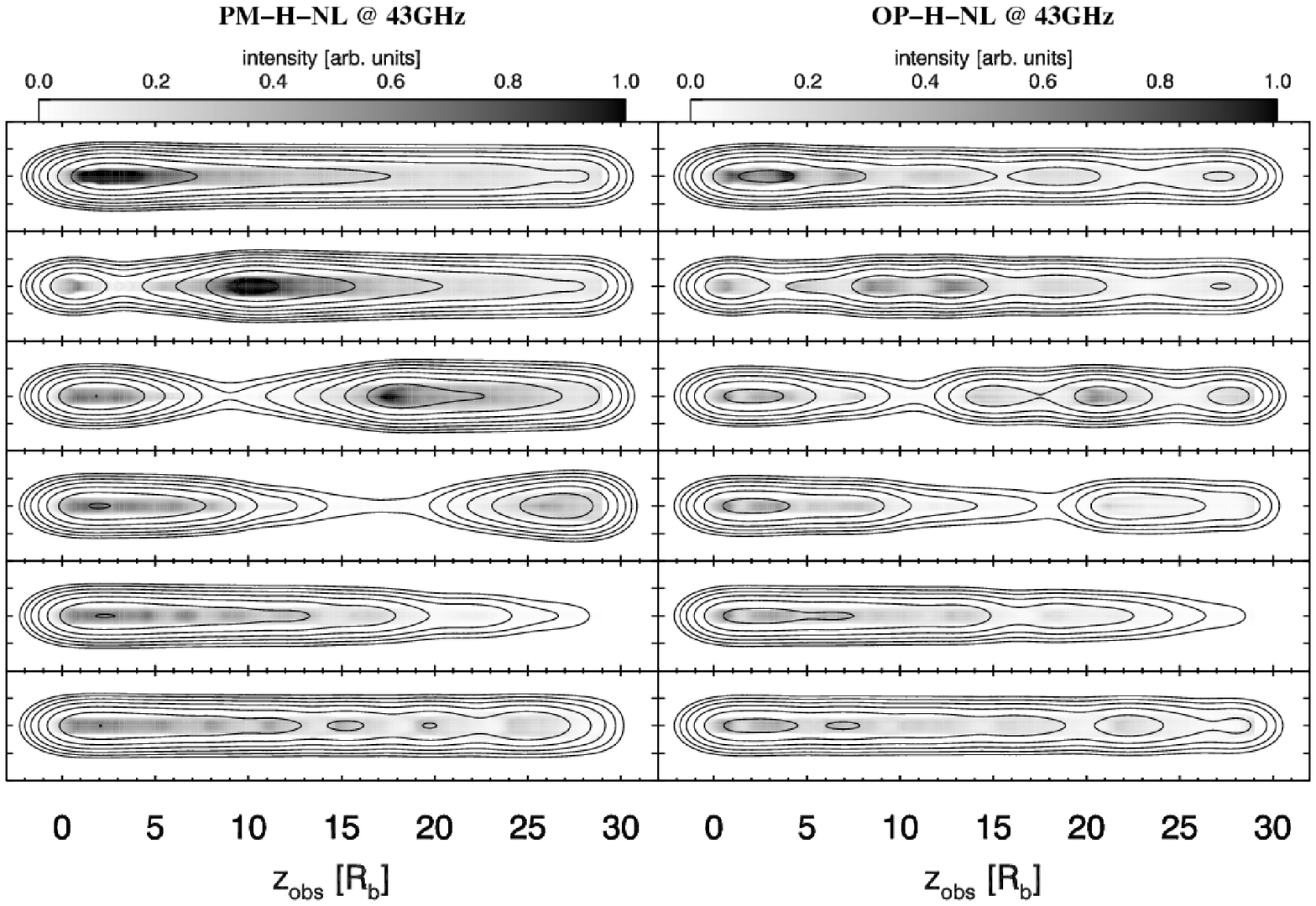}
\caption{Same as Fig.~\ref{fig:PMOP-highB-snap}, but {\cblue without
    including synchrotron losses (SPEV-NL method).}
  \label{fig:PMOP-highB-nl}}
\end{figure*}

\subsection{On the relevance of synchrotron losses}
\label{sec:losses}

The magnetic field energy density is set {\it ad hoc} in our models
(Sect.~\ref{sec:synchrotron}), and we can change it freely if the
resulting magnetic field does not become dynamically
relevant. 
\begin{figure*}
\plotone{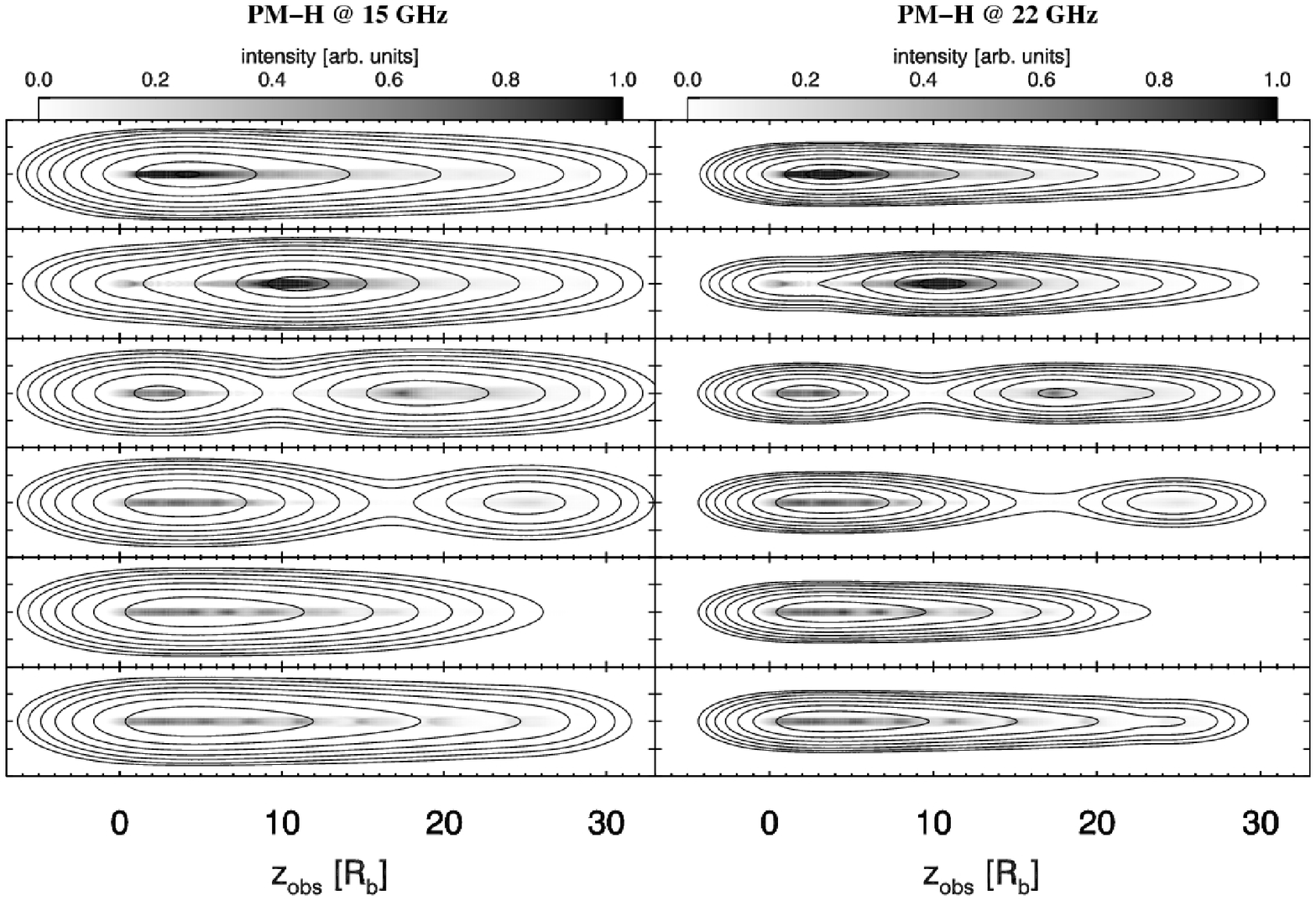}
\caption{Same as Fig.~\ref{fig:PMOP-highB-snap}, but only showing the
  {\cblue PM-H} model observed at 15\,GHz (left) and 22\,GHz (right). The images
  at each frequency use an intensity gray scale separately normalized
  to the maximum at the corresponding frequency. The vertical scale
  spans $20R_b$, i.e., there is a factor of two difference between the
  vertical scales shown in Fig.~\ref{fig:PMOP-highB-snap} and in this
  figure. Since the FWHM of the convolution beam depends linearly on
  the wavelength of observation, at 15\,GHz and 22\,GHz the FWHM are
  $6.45R_b$ and $4.40R_b$, respectively.   
  \label{fig:PM-highB-spectral-snap}}
\end{figure*} {\cblue For the sake of a better illustration of the
  effect of the synchrotron losses on the morphologies displayed in
  the radio maps, we have computed models PM-H and OP-H
  (Fig.~\ref{fig:PMOP-highB-snap}), and PM-H-NL and OP-H-NL
  (Fig.~\ref{fig:PMOP-highB-nl})}. A noticeable general characteristic
of SPEV-NL models is that all the features identifiable in the radio
maps are more elongated (along the jet axial direction) than in the
case that synchrotron losses are included. The reason is that without
synchrotron losses, the beam of the jet is brighter at longer
distances. Thus, in the unconvolved data, the parts {\cblue located
  downstream the jet weight more in the convolution beam} than in the
case where synchrotron losses are included, biasing the isocontours of
flux density along the axial, downstream jet direction. For the same
reason, the models which include synchrotron losses display a more
knotty morphology than those which do not include them, both in the
unconvolved and in the convolved data. This feature is more important
in case of {\cblue OP-H and OP-H-NL models} (compare, e.g., panels
two, three and six -from top- of Figs.~\ref{fig:PMOP-highB-snap} and
\ref{fig:PMOP-highB-nl}) than in case of {\cblue PM-H and PM-H-NL
  models}.

The {\cblue main component undergoes losses-dominated (first) and
  adiabatic (later) regimes as quiescent jet models do}.  In the
losses-dominated regime (upper two panels of
Fig.~\ref{fig:PMOP-highB-snap}), SPEV models exhibit a brighter
component than SPEV-NL models. Later, in the adiabatic regime, SPEV
models display a dimmer component than SPEV-NL ones. As we argued in
Sect.~\ref{sec:magfield}, the conservation of the NTPs number density
explains this phenomenology.

The main component clearly splits into two parts when synchrotron
losses are included in model OP-H (Fig.~\ref{fig:PMOP-highB-snap}
panels 2 and 3 from top; see also the movie ``PMOP-highB.mpg'' in the
online material). The component splitting is not so apparent in model
{\cblue OP-H-NL}, although it also takes place farther away from the
nozzle than in the model including losses
(Fig.~\ref{fig:PMOP-highB-nl}, third panel from top). The splitting of
the main component happens during the losses-dominated regime and the
rear part of the component is brighter than the forward one if losses
are included, otherwise, the forward part of the component is brighter
than the rear one. However, the fact that the component is seen as a
double peaked structure is not the direct result of the splitting of
the hydrodynamic perturbation in two parts (\S~\ref{sec:perturb}),
because the projected separation of these two hydrodynamic features is
smaller than the convolution beam, even at 43\,GHz. Instead, this
results from the interaction of the hydrodynamic perturbation with the
cross shocks in the beam of model OP. Because of the small viewing
angle, the increased emission triggered in the component when it
crosses over a recollimation shock is seen by the observer to arrive
simultaneously with the radiation emitted when the hydrodynamic
perturbation was crossing over the preceding (upstream) cross shock.

Figure~\ref{fig:PM-highB-spectral-snap} shows the evolution of the
component at 15\,Ghz (left panels) and 22\,GHz (right panels) for the
{\cblue PM-H model}. The convolution beam depends linearly on the
wavelength of observation, thereby, it is larger at smaller
frequencies.
Except for the obvious disparity of resolutions 
the evolution of the main component along the pressure matched jet at
15\,GHz, 22\,GHz and 43\,GHz {\cblue does} not display large
differences. The main component appears as a moving bright spot at all
three frequencies (upper three panels of
Fig.~\ref{fig:PMOP-highB-snap}~{\it left} and
Fig.~\ref{fig:PM-highB-spectral-snap}).
%
%
\begin{figure}
\plotone{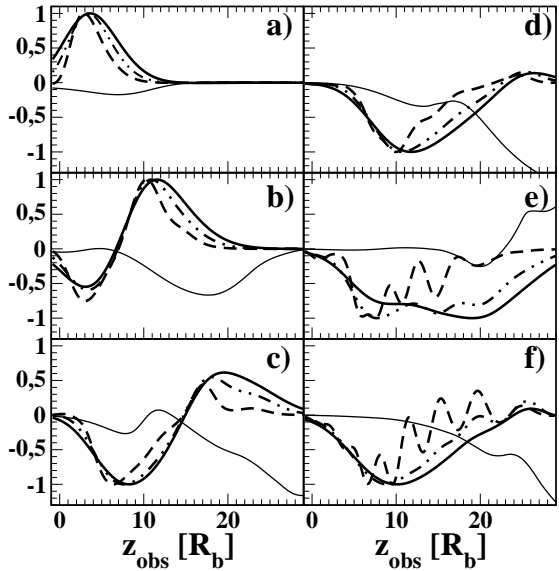}
\caption{In each panel lines correspond to the differences in
    intensity or spectral index between model {\cblue PM-H} with and without a
  hydrodynamic perturbation. Each panel
  corresponds to a different observer's time (the times are the same
  as in Figs.~\ref{fig:PMOP-highB-snap} and
  \ref{fig:PM-highB-spectral-snap}). The thick-solid, dashed-dotted
  and dashed lines represent the normalized difference $(I^p(Z) -
  I(Z)) / \max_{_{Z}}|I^p(Z) - I(Z)|$, at 15\,GHz, 22\,GHz and
  43\,GHz, respectively. $I(Z)$ and $I^p(Z)$ are the intensities
  averaged over cross-sections of the jet at each distance $Z$ from
  the injection nozzle { for the quiescent and perturbed models,
    respectively}. The maximum in the denominator extends for all $Z$
    along the jet axis. The thin solid line shows the difference in
    the spectral index between the {\cblue PM-H} model with the hydrodynamic
    perturbation and the corresponding quiescent model. Precisely, the
    line shows the function $5\times(\alpha_{13}(Z) -
    \alpha_{13}^p(Z))/\max_{_{Z}}|\alpha_{13}(Z)|$, where
    $\alpha^p_{13}(Z)$ and $\alpha_{13}(Z)$ correspond to the
    cross-sectional average of the spectral index
    (Eq.~\ref{eq:alpha_ij}) of the jet with the injected hydrodynamic
    perturbation and to the quiescent jet, respectively.
    \label{fig:PM-highB-spec}}
\end{figure}

\begin{figure}[hbt]
\plotone{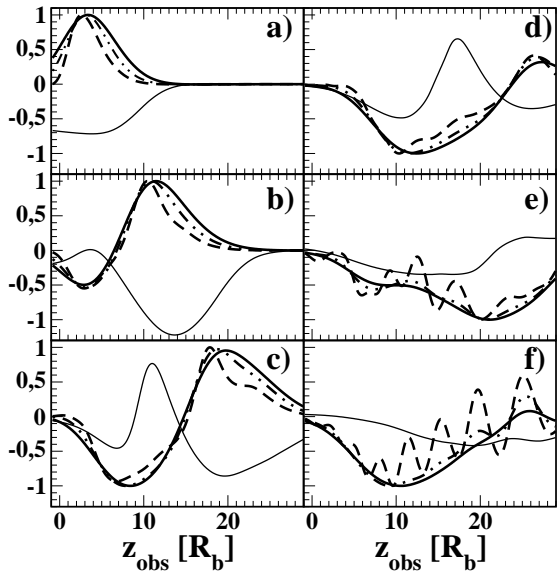}
\caption{Same as Fig.\ref{fig:PM-highB-spec} but for {\cblue the model PM-L}.
  \label{fig:PM-fiduc-spec}}
\end{figure}

We have also checked that the profile outlined above does not depend
on including synchrotron losses either. However, the smaller the
magnetic field, the larger the increase in the spectral index behind
the intensity maxima associated to the main component (i.e.,
associated with the rarefaction trailing the main hydrodynamic
perturbation). The time evolution of the prototype spectral profile of
a hydrodynamic perturbation injected at the nozzle is characterized by
a substantial steepening of the spectrum behind the intensity maxima
(Figs.~\ref{fig:PM-highB-spec}c and ~\ref{fig:PM-fiduc-spec}c,d)
compared to the quiescent jet model. This behavior of the spectral
index has also been found in previous theoretical papers, and it is
attributed to the fact that the NTP distribution evolves on timescales
smaller than the light crossing time of the source (e.g.,
\citealp{CG99}).

\begin{figure}
\plotone{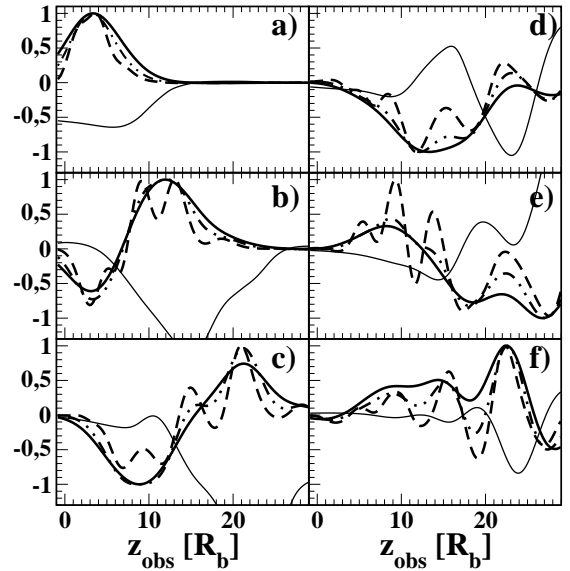}
\caption{Same as Fig.\ref{fig:PM-highB-spec} but for {\cblue the model OP-L}.
  \label{fig:OP-fiduc-spec}}
\end{figure}

Comparing Figs.~\ref{fig:PM-highB-spec}e and
~\ref{fig:PM-fiduc-spec}e, it is remarkable that trailing components
pop up precisely to the left (i.e., behind) of the local relative
spectral index minimum (at $Z\sim 18R_b$ in
Fig.~\ref{fig:PM-fiduc-spec}e and $Z\sim 20R_b$ in
Fig.~\ref{fig:PM-highB-spec}e) that follows the local relative maximum
of the spectral index reached in the wake of the main
perturbation. Furthermore, we notice that the intensity relative to
the background jet of the trailing components identifiable at 43\,GHz,
depends on the strength of the initial magnetic field, in spite of the
fact that in our models the magnetic field is dynamically negligible.%
\footnote{According to \cite{MAM07}, the boundary separating magnetic
  fields dynamically relevant from those in which the magnetic field
  is dynamically negligible is around $U_{\rm b}\simeq 0.03P$. In our
  case, even for the model with the largest comoving magnetic field,
  we have $U_{\rm b}=0.01P$.}
At higher magnetic field strength the intensity of the trailing
components is lowered and, some of them are hardly visible (e.g., the
leading trailing at $\sim 25R_b$ is evident in
Fig.~\ref{fig:PM-fiduc-spec}f, while it is difficult to identify in
Fig.~\ref{fig:PM-highB-spec}f). {\cblue Thereby,} the observational
imprint of trailing components is frequency dependent.

The evolution of the perturbation in model {\cblue OP-L} displays a
slightly different profile at 43\,GHz than in model {\cblue PM-L. The
  main component splits into two sub-components at the highest
  observing frequency (Fig.~\ref{fig:OP-fiduc-spec}b)}. At 15\,GHz and
22\,GHz, the profile of the perturbation is qualitatively the same as
for the {\cblue PM-L} model. {\cblue The} spectral index displays a
behavior very similar to that of the {\cblue PM-L} model. However, the
evolution after the passage of the main component in model {\cblue
  OP-L} (Fig.~\ref{fig:OP-fiduc-spec}d~--~\ref{fig:OP-fiduc-spec}f) is
different from that of model {\cblue PM-L}. The number of bright spots
popping up in the wake of the main perturbation is smaller and they
are brighter (in relation to the quiescent jet) in the {\cblue OP-L}
model than in the {\cblue PM-L} one. {\cblue Identifying} these
features as trailing components ({\cblue
  Sect.~\ref{sec:spacetime_analysis}}), we realize that they do not
only appear at 43\,GHz, but also at 22\,GHz, and one may guess them
even at 15\,GHz.

\subsection{Spacetime analysis}
\label{sec:spacetime_analysis}

\begin{figure*}
\plotone{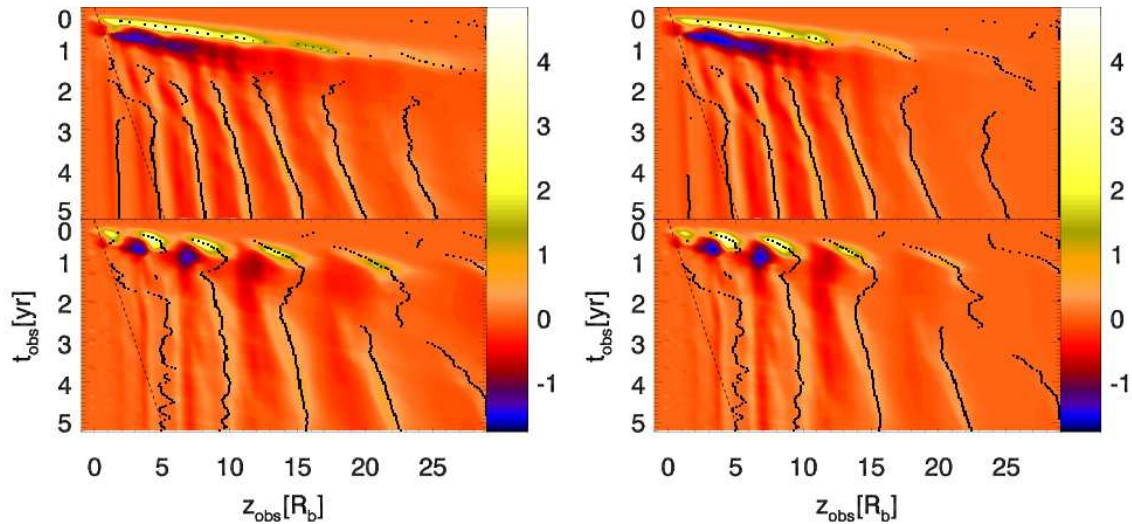}
\caption{Spacetime plot of the difference in the intensity at 43\,GHz,
  averaged over the beam cross-section, between the perturbed and
  quiescent SPEV emission for {\cblue PM-L and OP-L (upper left and lower
    left panels, respectively) and PM-H and OP-H (upper right and
    lower right panels, respectively)} unconvolved models.  The slope
  of the dashed line corresponds to an apparent velocity equal to the
  speed of light. The solid black dots correspond to the world-lines
  of a number of bright features observed in the convolved
  43\,GHz-radio images resulting from the difference between the
  hydrodynamic models with and without an injected perturbation. Among
  these features, there are trailing components (particularly in the
  {\cblue PM} models) and also standing recollimation shocks
  (characteristic of the {\cblue OP} models). The peaks in the color plot
  (yellow-white shades) do not always match the distribution of black
  dots due to the difference in the resolution of the convolved and
  unconvolved data. The color scale is linear and common for each
  column of panels. It displays the difference of averaged intensities
  in arbitrary units.
  \label{fig:PMOP-sptm}}
\end{figure*}

In order to relate the hydrodynamic evolution with the features
observed in the synthetic radio maps, we have built up several
space-time diagrams of the evolution of the component as seen by a
distant observer. In Fig.~\ref{fig:PMOP-sptm} we plot the difference
in intensity, averaged over the beam cross-section, between the
perturbed and quiescent models. This difference accounts for the net
effects that the passage of the hydrodynamic perturbation triggers on
the quiescent jet. The trajectory of the main component is seen as a
bright (yellow) region close to the top of each plot. {\cblue Its
  superluminal motion is apparent when the slope of the trajectory is
  compared to that of the dashed line, which denotes the slope
  corresponding to the speed of light. Below the main component, the
  dark (blue) region is associated to the reduced intensity that the
  rarefaction trailing the hydrodynamic perturbation leaves.

  As in G97, while in models PM-L and PM-H the main component and the
  reduced intensity region trailing it are continuous in the
  space-time diagrams (Fig.~\ref{fig:PMOP-sptm} upper panels), in OP-L
  and OP-H models they flash intermittently as they cross over
  standing cross shocks of the beam (larger intensity --
  Fig.~\ref{fig:PMOP-sptm} lower left panel) and then expand in the
  rarefactions that follow such standing shocks (smaller
  intensity). The interaction of the perturbation with the standing
  shocks of the quiescent OP model results in a displacement of the
  position of the shocks also noticed in G97. The temporarily dragging
  of standing components, is clearly visible in the lower left panel
  of Fig.~\ref{fig:PMOP-sptm}. The second (from the left) of the well
  identified bright spots, oscillates with an amplitude of $\sim 1.4
  R_b$ in $\sim 10\,$months. The trend being to increase both the
  oscillation period and the amplitude with the distance to the jet
  nozzle.

  Besides} the main component, we observe several trailing components
\citep{Agudoetal01}, identified in Fig.~\ref{fig:PMOP-sptm} by
``threads'' with an intensity larger than in the quiescent model,
which emerge from the wake of the main component. In
Fig.~\ref{fig:PMOP-sptm} we also overplot (black dots) the world-lines
of a number of bright features observed in the convolved 43\,GHz-radio
images resulting from the difference between the hydrodynamic models
with and without an injected perturbation. These world-lines show only
those local intensity maxima which could be unambiguously tracked in
{\cblue convolved radio maps}. Except for the bright features closer
to the jet nozzle, the world-lines match fairly well the unconvolved
trails of high intensity. 
The latest three trailing components of Fig.~\ref{fig:PMOP-sptm}
(upper left panel) do actually recede\footnote{\cblue Trailing
  components are pattern motions in the jet beam.} in the convolved
43\,GHz maps as much as $0.5R_b$ for 1 to 4 moths, soon after they are
identified (i.e., at an apparent speed $\sim 0.5c - 0.9c$).
\begin{figure*}
\plotone{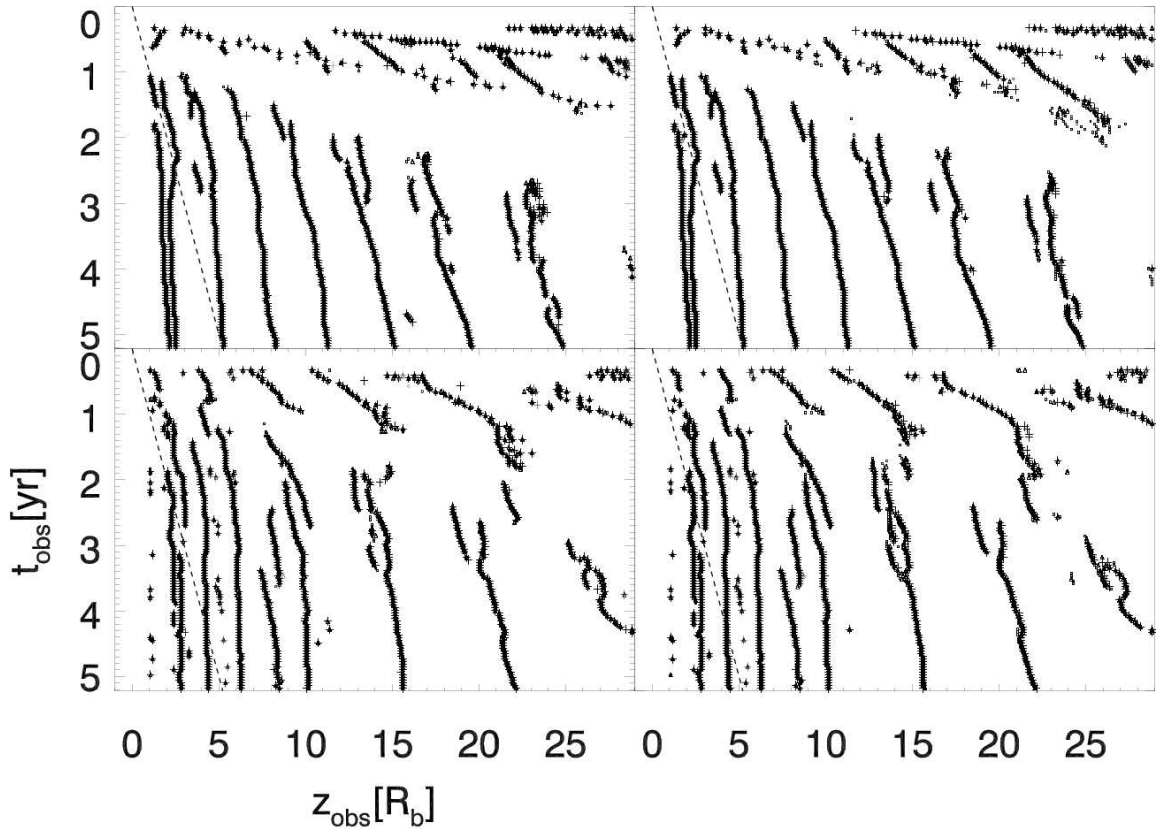}
\caption{World-lines of a number o bright features observed in
  the unconvolved radio images resulting from the difference between the
  hydrodynamic models with and without an injected perturbation (like
  in Fig.~\ref{fig:PMOP-sptm}). Squares, triangles and crosses
  correspond to the radio frequencies 43\,GHz, 22\,GHz and 15\,GHz,
  respectively. The size of the symbols is proportional to the
  wavelength of the data they display. Shown are {\cblue PM-L and OP-L (upper left and lower
    left panels, respectively) and PM-H and OP-H (upper right and
    lower right panels, respectively) models}. The slope of the dashed line
  corresponds to the an apparent velocity equal to the speed of light.
  \label{fig:PMOP-sptm-traj-nconv}}
\end{figure*}

{\cblue Like} in the case of PM models, in the wake of the main component
of model OP a number of bright spots seem to emerge with increasingly
larger apparent velocities as they pop up far away from the jet
nozzle. However, looking at the locations from where these components
seem to emerge, we notice that they are in clear association with the
locus of the standing shocks of the OP models. Such an association is
even more evident when we look at the world-lines of the brighter
features trailing the main component as they are localized in the
43\,GHz radio maps. {\cblue The} physical origin of these trailing
features differs from that of the trailing components seen in PM
models. {\cblue There} trailing components are local increments of the
pressure and of the rest-mass density of the flow produced by the {\it
  linear} growth of KH modes in the beam, generated by the passage of
the main hydrodynamic perturbation. In the beam of OP models,
intrinsically {\it non-linear} standing shocks are
present. Nonetheless, the interaction of a non-linear hydrodynamic
perturbation with non-linear cross shocks yields an observational
trace which resembles much that of a trailing component. Thereby we
keep calling such features trailing components, following
\cite{Agudoetal01}. 

If the jet is not pressure matched, all the KH modes excited in the
beam are blended with standing knots. Indeed, we realize that close to
the jet nozzle, the locus of the first two bright spots is almost
standing and, at larger distances, the subsequent knots show a clear
increment of its pattern speed. The fist two trailing components are,
actually, the traces of standing shocks which are dragged along with
the main perturbation and oscillate around their equilibrium
positions. The remaining trailing components move much faster and they
can probably be due to the pattern motion of KH modes in the OP beam.

Comparing the traces left by the passage of the main hydrodynamic
perturbation in the PM and OP models (Fig.~\ref{fig:PMOP-sptm}), it
turns out that the signatures of such perturbation are much cleaner
and numerous in {\cblue PM than in OP models}. The number of trailing
components is smaller in {\cblue OP than in PM models}, and their
world-lines are more oscillatory than in the latter case. {\cblue For a
  lager magnetic field (models PM-H and OP-H; Fig.~\ref{fig:PMOP-sptm}
  right panels)} NTPs cool faster and radiate more energy, {\cblue and
  thus,} one can basically see only features happening close to the
jet nozzle.

\begin{figure*}[hbt]
\plotone{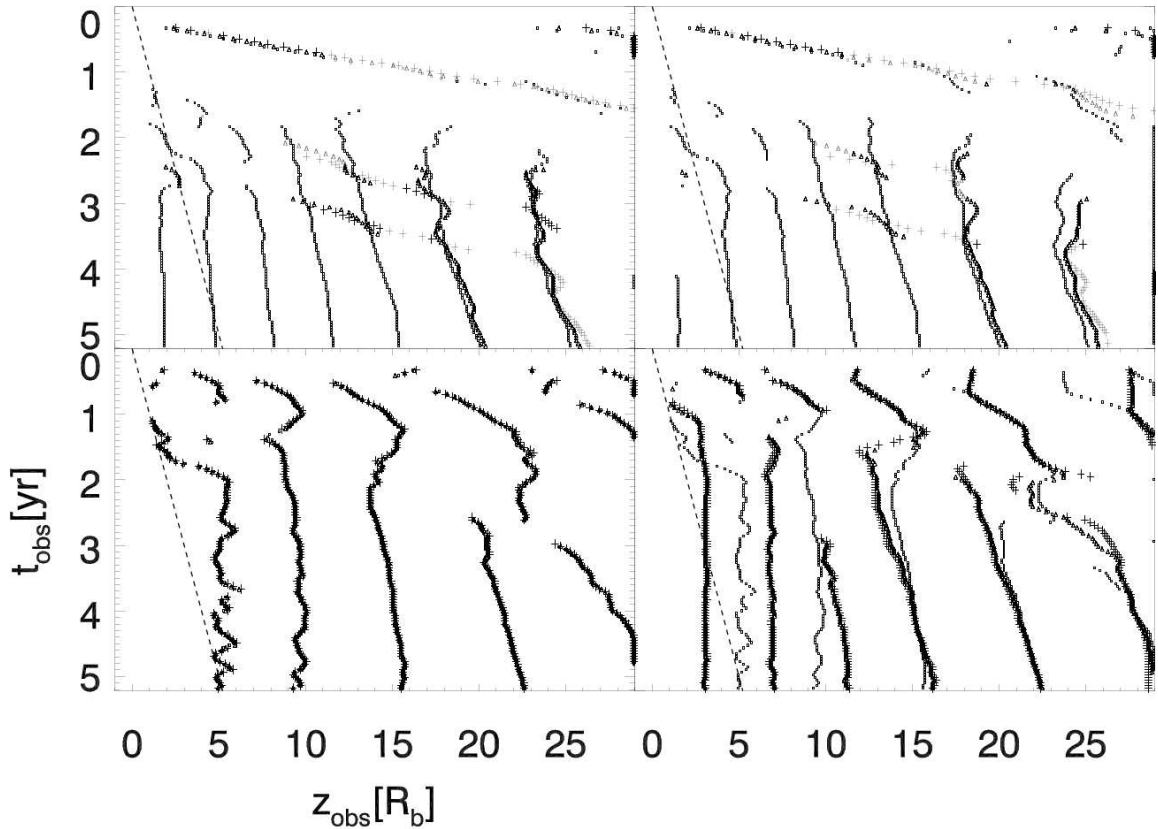}
\caption{Same as Fig.~\ref{fig:PMOP-sptm-traj-nconv} but using
  convolved data at every frequency. The location of local
    maxima trailing the main perturbation in the beam of PM models is
    difficult at 15\,GHz and 22\,GHz because of the large convolution
    beam at these two frequencies. The low observing resolution at
    such frequencies drives spurious detections of local maxima in the
    radio maps (trailing components), which explain the anomalous data
    at 15\,GHz and 22\,GHz in the range $(t_{\rm obs}, z_{obs}) =
    (2\,{\rm yr}- 3.7\,{\rm yr}, 10R_b -20R_b)$ in the PM model (upper
    panels). The OP models do not display such obvious anomalies
    because the high intensity threads that trail the main
    perturbation are associated with preexisting recollimation shocks
    in the beam, whose emissivity, relative to the background jet, is
    much larger than that corresponding to the trailing components in
    the PM models.
    \label{fig:PMOP-sptm-traj-conv}}
\end{figure*}

%
%
The unconvolved data for both PM and OP models, and independently of
the magnetic field strength, is compatible with not having any time
lag between the high and low frequency radiation emitted by the main
component, i.e., the radiation at all three frequencies is co-spatial
(Fig.~\ref{fig:PMOP-sptm-traj-nconv}). However, the convolved data
display a number of positive and negative time lags which result from
the difference in the size of the convolution beam at every
frequency. In case of the PM models, there is a trend of the 43\,GHz
maximum emission to lie behind the corresponding maxima at 22\,GHz and
15\,GHz (Fig.~\ref{fig:PMOP-sptm-traj-conv} upper panels). Thereby,
the low energy radiation from the main component is seen first, and
later an observer detects radiation at higher frequencies.
Nevertheless, considering that the resolution of the convolved data is
worse at smaller frequencies, the emission from the component is
consistent with having no time-lags between low and high frequency
emission. This trend is independent of the magnetic field strength,
but it is more obvious for the model {\cblue PM-H model} (note the large
separation between the different symbols beyond $Z_{\rm obs} \sim 15
R_b$ in the Fig.~\ref{fig:PMOP-sptm-traj-conv} upper right
panel). Therefore, any positive or negative time lag of radiation at
different frequencies measured from convolved data has to be taken
with care.

For OP-L models, positive and negative time lags between the high and
low energy radiation are observed along the $z$-axis
(Fig.~\ref{fig:PMOP-sptm-traj-conv} lower left panel). Such time lags
are smaller than for the {\cblue PM-H} model and, indeed, the data are
compatible with no-time lags at all. For {\cblue OP-H}, in most cases,
the high-frequency emission dots lie in front of the lower frequency
ones (Fig.~\ref{fig:PMOP-sptm-traj-conv} lower right panel). But
still, considering the difference in linear resolution for the
location of the maxima, the radiation at different frequencies is
almost co-spatial.

%

%
%
Trailing components can only be tracked at 43\,GHz close to the jet
nozzle. Only after a certain distance, it is possible to see them at
22\,GHz and even at 15\,GHz (see the last two trailing world lines in
each panel of Fig.~\ref{fig:PMOP-sptm-traj-conv}). The world lines of
trailing components at 22\,GHz and, particularly, at 15\,GHz, undergo
substantial velocity changes. During some time intervals the convolved
data shows receding trailing components at such frequencies. In the OP
models, there are no clear trends, independent of the magnetic field
strength, since it is very difficult to locate any local maxima at
15\,GHz, and the 22\,GHz data lie almost on top of the 43\,GHz
points. We note that there is a mismatch between the data points at
different frequencies in the {\cblue OP-H model} at the first two recollimation
shocks (vertical threads at $z_{\rm obs}\sim 4R_b$ and $7R_b$). It is
produced because there is a rather small relative difference in the
emissivity of the perturbed and the quiescent jet models at 43\,GHz
until $z_{\rm obs} \lsim 10R_b$. In such conditions, the algorithm to
detect local maxima in the space-time diagrams yields oscillatory
results. A large mismatch between the world-lines of the peak
intensity of trailing components at different frequencies also happens
in other trailing features (e.g., the fourth and fifth threads in
Fig.~\ref{fig:PMOP-sptm-traj-conv} lower right panel). This mismatch
does not exist in the corresponding unconvolved data
(Fig.~\ref{fig:PMOP-sptm-traj-nconv}) and, hence, we conclude it is an
artifact of the finite size of the convolution beam at the observing
frequencies.

\section{Discussion and conclusions}
\label{sec:conclusions}

%
%
We have presented a new method (SPEV) to compute the evolution of NTPs
coupled to relativistic plasmas under the assumption that these NTPs
do not diffuse across the underlying hydrodynamic fluid. NTPs change
their energy because of the variable hydrodynamic conditions in the
flow and because of their synchrotron losses in an assumed background
magnetic field. The inclusion of synchrotron losses and a transport
algorithm for NTPs are major steps forward with respect to previous
approaches we have followed. The new method has been validated with
another preexisting algorithm suited for the same purpose, but without
including synchrotron losses and transport of NTPs (AM algorithm). The
validation process shows that the SPEV method reproduces the same
qualitative phenomenology as outlined in the previous works of our
group (G95, G97). The power of the new method in its whole blossom
shows up when synchrotron cooling dominates the NTP evolution.

%
%
\paragraph{Quiescent jet models:} When synchrotron losses are
considered, the resulting phenomenology can be split into two regimes:
losses-dominated and adiabatic regime (following the convention of
\citealp{MG85}). In the losses-dominated regime, the knots displayed
in the radio maps, which are close to the jet nozzle, are brighter
than in models which do not include synchrotron cooling at the
considered frequencies. {\cblue Indeed, quiescent jet models including
  radiative losses are more knotty than those models which do not
  include them. These features result} from the conservation of the
number density of NTPs. Since the same number of particles per unit of
volume that initially extends from $\gamma_{\rm min}(t=0)$ to a
certain upper limit $\gamma_{\rm max}(t=0)$ is confined into a
narrower Lorentz factor interval, wherein more NTPs are efficiently
emitting in the considered observational radio bands. In the adiabatic
regime (reached relatively far away from the jet nozzle), the spectral
changes, that the NTP population experiences as it is advected
downstream the jet, of models with and without losses is qualitatively
similar, since most of the high-energy NTPs (which evolve faster) have
cooled down to energies where losses are negligible. The beam of the
jet in the adiabatic regime is dimmer at radio-frequencies than in
models where synchrotron losses are not included. Our method lacks of
a suitable scheme to account for diffusive shock acceleration of
NTPs. However, all shocks existing in the quiescent jet models are
rather weak and, for practical purposes they can be considered as
compressions in the flow, where an enhanced emission is obtained due
to the local increase of density and of pressure.

{\cblue One of the main results of this work is that for the same
  background hydrodynamic jet model, dynamically negligible magnetic
  fields of different strengths yield substantially different observed
  morphologies. This introduces a new source of degeneracy (in
  addition to relativistic effects, such as, time delay, aberration,
  etc.) when inferring physical parameters out of observations of
  radio jets. For example, the difference in the observational
  properties of models OP-L and OP-H (Sect.~\ref{sec:synlos}) shows,
  that increasing the magnetic field strength by a factor of 10
  triggers a much faster cooling of the NTPs, resulting in a much
  shorter losses-dominated regime and shorter jets, despite magnetic
  field remaining dynamically unimportant.} Furthermore, jet models
with such a large magnetic field display a larger flux density
contrast between shocked and unshocked jet regions. The reason being
that after the losses-dominated regime, $\gamma_{\rm max}$ is reduced
so much that most of the NTP population is inefficiently radiating at
the considered radio wavelengths and, only when the non-thermal
electrons are compressed at cross shocks of the beam, they partly
reenter into the efficiently radiating regime at the considered
frequencies.

%
%
\paragraph{Spectral inversion:} {\cblue In this paper we suggest that an
  inverted spectrum may also result if the lower limit of the NTP
  distribution $\gamma_{\rm min}$ is larger than the value of
  $\gamma_{_{\rm M}}$ for which the synchrotron function $R(x)$
  reaches its maximum (Eq.~\ref{eq:gamma_min2}), in agreement with the
  theoretical predictions of \cite{TK07}.}  Evidences for flat,
optically thin radio spectra in several active galactic nuclei have
been shown by, e.g., \cite{HAA89b,Melrose96}, and \cite{Wangetal97}.
These authors consider different kinds of Fermi-like acceleration
schemes to be responsible for the hardness of the electron energy
spectra. \cite{SP08} show that stochastic interactions of radiating
ultrarelativistic electrons with turbulence characterized by a
power-law spectrum naturally result in a very hard (actually inverted)
electron energy distribution which yields a synchrotron emissivity at
low frequencies with an spectral index $\simeq 1/3$. Alternatively,
Birk, Crusius-W\"atzel \& Lesch (2001) argue that optically thin
synchrotron emission due to hard electron spectra produced in magnetic
reconnection regions may explain the origin of flat or even inverted
spectrum radio sources. {\cblue In contrast to our findings, these
  authors, explain the spectral inversion in some sources as a result
  of a flatter electron energy distribution. Observationally, it could
  be possible to discriminate between both possibilities by looking at
  the high-energy spectrum of the source. If there are external seed
  photons (e.g., from the AGN), which were Compton up-scattered by the
  non-thermal electrons of the jet, the spectral index at high
  energies could discriminate between the alternative explanations for
  the optically-thin inverted spectra at radio frequencies.}

Since $\gamma_{\rm min}$ is fixed in our model through
Eq.~(\ref{eq:pmin}) and it is not derived from first principles, one
may question whether the value we obtain for $\gamma_{\rm min}$ could
be too large and, therefore, the spectral inversion we are explaining
on the basis of taking $\gamma_{\rm min}\gsim\gamma_{_{\rm M}}$ is
unlikely to happen in nature. This would be the case if the jet was
composed by and electron-positron plasma, in which case $\gamma_{\rm
  min}\simeq 1$ (e.g.,\citealt{Marscheretal07}). For plasmas made out
of electrons and protons, \cite{Wardle77} obtained that for
synchrotron sources with a brightness temperature $\simeq 10^{12}\,$K
and $q=2$, $\gamma_{\rm min}\gsim 161$ in order to account for the low
degree of depolarization in parsec-scale emission regions. More
recently, \cite{Blundelletal06} inferred $\gamma_{\rm min}\sim 10^4$
at the hot-spots of 6C\,0905+3955 (see also \citealt{TK07}, and
references therein). Thus, the exact value of $\gamma_{\rm min}$ is
probably source dependent, and our minimum Lorentz factor threshold
($\gamma_{\rm min}\simeq 330$) can be well accounted by present day
theory and observations if the jet is not a pure electron-positron
plasma.

%
%
\paragraph{Radio components:}
We have applied the SPEV method to calculate the spectral evolution of
superluminal components in relativistic, parsec-scale jets. These
components are set up as hydrodynamic perturbations {\cblue at the jet
  nozzle}. For a small value of the magnetic field (the same as in
G97), synchrotron losses are negligible and we recover the
phenomenology shown by G97 and \cite{Agudoetal01}. 

{\cblue The main component is characterized by a hardening of the
  spectrum. Pressure matched models yield a generic spectral profile
  of the component, which is rather independent of synchrotron
  losses. The hydrodynamic perturbation looks in the radio maps like a
  burst at every radio-frequency and, just behind it, there is a
  decrease of the flux density. The shape of the burst is asymmetric
  in the axial jet direction, being brighter upstream than
  downstream. The shape of the burst is also frequency dependent
  because the convolution beam grows linearly with the observing
  wavelength (at lower frequencies the component is more symmetric in
  the axial jet direction). This triggers a decrease of the spectral
  index in the forward region of the main component, until it reaches
  a minimum (which precedes the intensity maxima at the highest
  observing frequency).

When radiative losses are important, a number of differences can be
observed:
\begin{enumerate}
\item Main component splitting in OP-H model: The main component
  splits in the radio-maps much more clearly than in OP-L model
  (Sect.~\ref{sec:losses}), and the splitting takes place farther away
  from the nozzle in the latter than in the former case. The rear part
  of the component is brighter than the forward one if losses are
  included. The spectral index profile is unaffected by the apparent
  splitting of the component. We conclude that the apparent splitting
  of the main component is an artifact of the sampling of the results
  in the observer frame. It is necessary to perform a finer time
  sampling of the radio jet than the $\sim 3.5\,$months we have
  considered in the radio maps, in which case, the main component
  exhibits an intermittent variation of its flux density (see on line
  material).  If observations do not have the sufficient time
  resolution, there is another hint that can help to disentangle
  whether the splitting is apparent or real. In a true splitting of
  the component, each part may show a different spectral aging due to
  their different hydrodynamic evolutions.
\item Radio features: Main and trailing components display a less
  elongated aspect in radio maps. The reason is that without losses,
  the beam itself is brighter at longer distances. Thus, in the
  unconvolved data, the parts located downstream the jet weight
  more in the convolution beam than in the case where synchrotron
  losses are included. For the same reason, models which include
  synchrotron losses display a more knotty morphology.  In the
  losses-dominated regime, SPEV models exhibit a brighter main
  superluminal component than SPEV-NL models. This behavior reverses
  in the adiabatic regime. Also, the ratio between the peak specific
  intensity of a trailing component to the specific intensity of the
  region of the beam immediately behind it, is larger than if losses
  are not included. The conservation of the NTPs number density
  explains this phenomenology (Sect.~\ref{sec:magfield}).
\item Spectral properties: Behind the main component, the spectral
  index returns almost monotonically to its unperturbed value. In
  contrast, when losses are negligible, there is a softening of the
  spectrum, just behind the main component (where the spectral index
  reaches a maximum).

\end{enumerate}}

%
%
\paragraph{Time lags:} {\cblue In this paper we explicitly show,}
that the convolved data has to be interpreted {\cblue
  carefully. During most of the time the main component is observable,
  the radiation emitted by the component at low energy (15\,GHz)
  arrives to the observer before that at high energy
  (43\,GHz). Indeed, for models PM-H and OP-H, a substantial mismatch
  between the world-lines of the peak intensity of trailing components
  at different frequencies is possible. This mismatch is an artifact
  due to the finite size of the convolution beam at the observing
  wavelengths.In contrast}, the unconvolved data is consistent with a
simultaneous emission of radiation at the three frequencies under
consideration. This behavior matches our expectations, since the
interval of observing wavelengths is too narrow to display a
substantial {\cblue frequency dependent} separation of the regions of
maximum emission.


%
%
\paragraph{On the nature of trailing components:} The journey of the
main component downstream the jet generates a number of frequency
dependent bright spots which pop up in its wake. {\cblue They
  differentiate themselves from the main component because (1) they do
  not emerge from the jet core, (2) they posses substantially smaller
  (sometimes subluminal or even, receding) speeds \citep{Agudoetal01}
  and, as we demonstrate here, (3) they do not exhibit an obvious
  change in the spectral index with respect to the quiescent jet
  model, but (4) their observational imprint is frequency dependent
  (they are clearly visible at the highest radio-observing
  frequencies, but at 22\,GHz and, particularly, at 15\,GHz they are
  wiped out by the large convolution beams at this wavelengths).} In
pressure matched jet models, {\cblue trailing components result from
  the linear growth of KH modes in the beam, after the passage of the
  main hydrodynamic perturbation \citep{Agudoetal01}. Here, we also
  consider overpressured jet models, where the situation is
  qualitatively different from pressure matched ones,} since the beam
of such models develops standing shocks ({\it non-linear}
structures). Nonetheless, the interaction of a non-linear hydrodynamic
perturbation with non-linear cross shocks yields an observational
trace which resembles that of a trailing component. Therefore,
sticking to the definition of \cite{Agudoetal01}, we also call
trailing components to the bright spots following the main component
in overpressured models, although the dynamical origin of such
components differs. {\cblue In this sense, every bright spot that
  results from the interaction between a strong hydrodynamic
  perturbation with a relativistic beam, which moves slower than the
  main component and is not ejected from the jet core shall be
  considered as a trailing component.} We shall add an obvious
cautionary note: striving for the knowledge of the jet parameters, on
the basis of a fit of the intensity variations behind a main
perturbation to a number or KH modes, requires that the jet is
pressure matched {\cblue (if the jet is not pressure matched, all the
  KH modes excited in the beam are blended with standing knots and the
  predicted jet parameters might be inaccurate). Furthermore,} it is
necessary that the linear resolution of the convolved (observational)
data was rather good. We have tested that the unconvolved results are
roughly recovered if the FWHM of the beam at 43\,GHz is smaller than
$0.25 R_b$. Insufficient linear resolution biases the observed
features in hardly predictable ways, rendering inadequate the
identification of features in the radio maps with hydrodynamic
structures.

In the future we plan to apply the SPEV method to perform additional
parametric studies of relativistic parsec scale jets. Among the
parameters which can be interesting to look at, we give preference to
the electron spectral index. Also, the SPEV algorithm can be coupled
to relativistic magnetohydrodynamic codes. This will drop any
assumption about the topology and strength of the magnetic field in
the jet, and it will enable us to perform also parametric studies of
polarization of the jet emission and of superluminal components.

\acknowledgments{P.~M. has performed this work 
with a European Union Marie Curie Incoming International
  Fellowship (MEIF-CT-2005-021603) and with the partial support
  obtained through the grant CSD-2007-00050. M.~A.~A. is a Ram\'on y
  Cajal Fellow of the Spanish Ministry of Education and Science. We
  acknowledge the support by the Spanish Ministerio de Educaci\'on y
  Ciencia and the European Fund for Regional Development through
  grants AYA2007-67626-C03-01, AYA2007-67626-C03-02 and
  AYA2007-67627-C03-03. I.~A. has been supported an I3P contract with
  the Spanish Consejo Superior de Investigaciones Cient\'{i}ficas. The
  authors thankfully acknowledge the computer resources technical expertise 
  and assistance provided by the Barcelona Supercomputing
  Center. We also thank the computer time obtained through the Spanish
  Supercomputing Network.}

\appendix

\section{A. Imaging algorithm}
\label{appendix:imaging}

Equations given in Secs.~\ref{sec:SPEV} and \ref{sec:synchrotron} are,
in principle, sufficient to compute the synchrotron emissivity at any
position in space and at any instant of time in the observer's frame,
either using SPEV or AM methods, accounting for the appropriate
transformations from the frame comoving with the fluid (where the
emissivity [Eq.~\ref{eq:j(nu)}], absorption coefficient
[Eq.~\ref{eq:kappa(nu)}], number density of NTPs [Eq.~\ref{eq:n^a}],
etc. are computed).{\cblue The purpose of this Appendix is to explain the
algorithm used to produce synthetic radio maps from discrete spatial
and temporal elements.}


\subsection{Geometry and arrival time}

While {\cblue in our simulations} the hydrodynamic state of the fluid
is axisymmetric regardless of the jet viewing angle, the observed
emission is, in general, not axisymmetric. We introduce the azimuthal
angle $\phi$ (measured in the $xy$-plane from the $x$-axis) and define
the laboratory frame {\cblue (attached to the center of the AGN)}
3D Cartesian coordinate system {\cblue $(x, y, z) := (R\cos\phi, R\sin\phi, Z)$, where the $z$-axis coincides with the jet
axis.}
%
%
We denote the jet viewing angle by $\theta$, and choose the following
observer coordinate system (rotated with respect to the 3D Cartesian
system by an angle $\theta$ around the $y$-axis)
\begin{equation}\label{eq:3Dobs}
  (x_{\rm obs},y_{\rm obs},z_{\rm obs}) := (x\cos\theta + z\sin\theta, y, -x\sin\theta + z\cos\theta)\,
\end{equation}
in which the observer is located along the $z_{\rm obs}$ axis, far
from the jet. For a given elapsed simulation time $T$ in the jet frame
the time of observation $t_{\rm obs}$ is defined as
\begin{equation}\label{eq:tobs}
  t_{\rm obs} := T - z_{\rm obs}/c
\end{equation}
The task of the imaging algorithm is to produce image in the $(x_{\rm
  obs}, y_{\rm obs})$ plane for a fixed arrival time $t_{\rm obs}$
(note that the image will be symmetric with respect to the $x_{\rm
  obs}$-axis if the magnetic field is completely random). From
Eqs.~\ref{eq:3Dobs}~-~\ref{eq:tobs} it is clear that we need to have
information about states of the jet at multiple instants of laboratory
frame time in order to correctly compute the contribution at a single
$t_{\rm obs}$. In a numerical hydrodynamic simulation we only have a
finite number of discrete iterations, but each iteration has an
associated time step $\Delta T$. In order to correctly take this into
account, in the following we assume that the time instant $t_{\rm
  obs}$ has a finite duration $\Delta T$ as well, and all radiation
arriving between $t_{\rm obs} - \Delta T/2$ and $t_{\rm obs} + \Delta
T/2$ is arriving precisely at $t_{\rm obs}$.

\subsection{Particle images}

Owing to the axisymmetric nature of the problem, we only follow the
Lagrangian particle motion and evolution in two dimensions (see
Sec.~\ref{sec:SPEV}). However, for the purposes of imaging, a
three-dimensional particle distribution needs to be created. We assume
that each particle which is injected at the jet nozzle has a radius
$\Delta r := R_{\rm b} / (2N_{\rm p})$, where $R_{\rm b}$ is the beam
radius and $N_{\rm p}$ number of particles per beam radius. That means
that a particle in two dimensions correspond to a revolution annulus
in the $(x, y, z)$ coordinate system\footnote{Note that also annuli
  are generated from the rotation of 2D cylindrical numerical cells
  around the jet axis and, thereby we can apply the same imaging
  procedure when we use the cell-based algorithm AM.}. In principle,
by knowing the particle position $(R_{\rm p}, Z_{\rm p})$ in the 2D
grid we could compute from Eqs.~\ref{eq:3Dobs}~-~\ref{eq:tobs} all
combinations of $(x, y, z)$ and, hence, all combinations $x_{\rm
  obs}$, $y_{\rm obs}$ and $t_{\rm obs}$ to which the particle annulus
corresponds for a fixed $T$. In practice, we approximate every annulus
by a series of cubes which are distributed along a circle with radius
$R_{\rm p}$, whose center is in $(0, 0, Z_{\rm p})$. The number of
cubes, evenly distributed in the azimuthal direction, necessary for an
optimal volume coverage of the annulus depends on the relation between
$R_{\rm p}$ and the particle radius $\Delta r$ (see next
subsection). By virtue of the symmetry of the jet, as seen by the
observer, with respect to the $x_{\rm obs}$-axis, we only need to
compute the contribution from one half-annulus, i.e. for those cubes
where $y = y_{\rm obs} \geq 0$.

{\cblue We} assume that both the emissivity and the absorption
coefficient are homogeneous within each cube. Thus, knowing the
particle velocity and the azimuthal angle of a given cube, we can
transform its emissivity and its absorption into the observer frame.

\begin{figure}
\plottwo{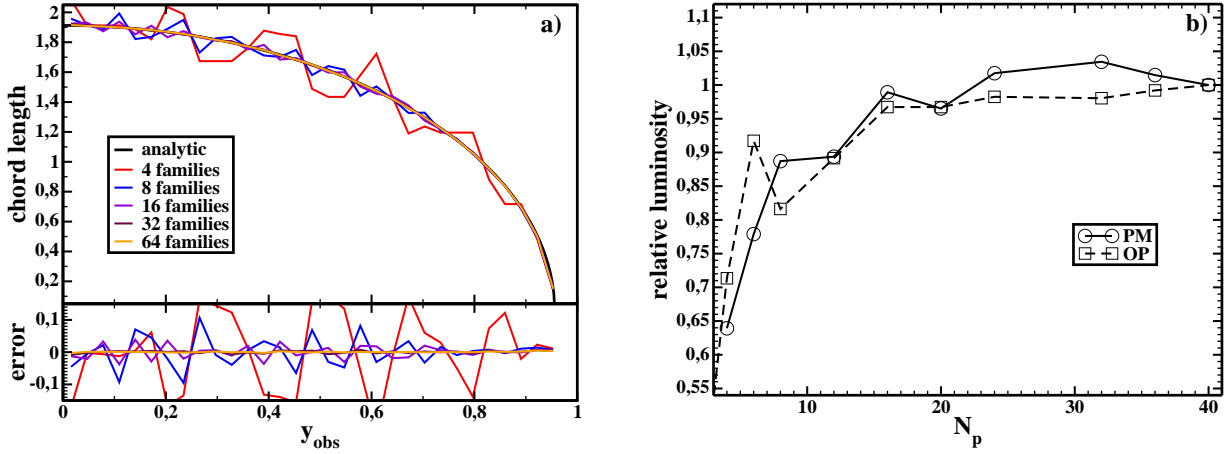}{FIGURES/fg20.eps}
\caption{{\it Left}: Results of the volume filling convergence test for different
  numbers of particles per unit of beam radius $N_{\rm p}$. In the
  upper panel we show the computed length of the chord through the jet
  as a function of the height $y_{\rm obs}$. We fix the value of the jet radius
  to be $R_{\rm b} = 0.95625$. The thick black line corresponds to the
  analytic expectation of the chord length, i.e., $2\sqrt{R_{rm b}^2 -
    y_{\rm obs}^2}$. In the lower panel the relative error with
  respect to the analytic expectation is displayed. {\it Right}: Results of the flux convergence test. The figure
  displays the total flux of images at $15\,$GHz for the
  quiescent PM (circles) and OP (squares) models as a function of the
  number of particles per beam radius $N_{\rm p}$ injected at the jet
  nozzle. Images have been produced with 4, 8, 16, 20, 24,
  32 and 40 particles per beam radius. The luminosity of both
  families of models are normalized to the total flux produced 
  by the corresponding (PM or OP) model with $N_{\rm p}=40$.
  \label{fig:circle}}
\end{figure}

\subsubsection{Approximation of annuli by cubes}

Given a particle with radius $\Delta r$ and cylindrical coordinates
$(R_{\rm p}, Z_{\rm p})$, we approximate the corresponding
half-revolution annulus by cubes evenly tessellating a circumference
centered at $(0, 0, Z_{\rm p})$. The angular separation between the
cubes is defined as $\Delta\phi = \min{(\pi, 2 \Delta r / R_{\rm
    p})}$. {\cblue Thus,} there are $N_\phi = \pi/\Delta \phi =
\max{(1, R_{\rm p} \pi / (2\Delta r))}$ cubes in a half-annulus
{\cblue of} volume $V_{\rm cubes}=N_\phi (2\Delta r)^3 = 4\pi R_{\rm
  p}(\Delta r)^2$ whereas the true volume of the half-annulus is
\begin{equation}
V_{_{\rm HA}} = \left\{
\begin{array}{rl}
[(R_{\rm p} + \Delta r)^2 - (R_{\rm p} - \Delta r)^2]\pi(2\Delta r)/2
= 4\pi R_{\rm p}(\Delta r)^2 & \rm{if\ } R_{\rm p}>\Delta r\\[4mm]
(R_{\rm p} + \Delta r)^2\pi(2\Delta r)/2 = (R_{\rm p} + \Delta
r)^2\pi\Delta r & \rm{if\ } R_{\rm p}\leq \Delta r
\end{array}
\right.
\end{equation}
{\cblue In the limit of $R_{\rm p} > \Delta r$,} the total volume of
the cubes is approximating that of the half-annulus. For $R_{\rm
  p}<\Delta r$, $N_\phi$ is reset to $1$ and the volume for which
$y>0$ is always $4(\Delta r)^3$, which is close to the average over
all possible values $R_{\rm p}\le \Delta r$ of the true volume
$7\pi(\Delta r)^3/6$.

\subsection{Radiative transfer}

{\cblue To compute an image
  we subdivide} the $(x_{\rm obs}, y_{\rm obs})$ plane into
rectangular pixels, and compute the contributions to each pixel by
checking which particle cube\footnote{One might also use spheres
  instead of cubes, but we use cubes to avoid dealing with
  trigonometric functions and square roots when checking for the
  intersection between rectangular pixels and particles.} intersects
which pixel at the right observation time. The ratio of the area of
intersection to the pixel area, gives a ``weight'' of the contribution
of a particular cube to the intensity {\cblue of the pixel}. For a
given $T$, the value of $z_{\rm obs}$ for each {\cblue particle} gives
the distance from the observer, so that we create a ``line-of-sight''
(LoS) for each pixel and sort along this line all contributing
particles according to {\cblue $z_{\rm obs}$} (note that these
contributions generally come from different instants of the laboratory
frame time $T$). Since in every pixel we sum up the contributions
spanning the observer time range $[t_{\rm obs}-\Delta T/2, t_{\rm
  obs}+\Delta T/2]$, the intersections of every LoS with particle
cubes are segments, not points (which would be the case if in every
pixel we would only consider the instantaneous contributions at
$t_{\rm obs}$). After all the contributions (i.e., intersection
segments) to a pixel have been accounted for, we solve the standard
radiative transfer equation to evaluate the final pixel intensity. The
above procedure can be performed simultaneously for a number of
different values of $t_{\rm obs}$, so that a ``movie'' in the observer
frame can be created. In order to transform the intensity detected in
a pixel into a flux we need to multiply by the pixel area.

\subsection{Tests of the method}

In order to validate our imaging algorithm we have developed two tests
which are based upon the idea that, increasing the number of Lagrangian
particles, both the volume filling factor%
\footnote{We define the volume filling factor as the fraction of the
  jet volume occupied by our finite size Lagrangian particles.}
and the total detected flux should converge. {\cblue We first show the
  convergence of the volume filling method. Then we show that the
  images and the total flux of the quiescent PM-L and OP-L models
  converge with increasing number of particle families.}

%
%

\subsubsection{Volume filling}

We have created a toy model consisting of a cylindrical {\em jet} with
uniform velocity parallel to the jet axis, and with a length equal to
the particle size $\Delta r$. The half-volume of such jet is $V_{\rm
  j,1/2}=\pi R_{\rm b}^2\Delta r$. We inject $N_{\rm p}$ particles in
the jet evenly distributed across the jet radius (i.e., $\Delta
r=R_{\rm b}/(2N_{\rm p})$, or $R_{\rm b}=(2i+1) \Delta r$, $i=0,
\dots, N_{\rm p} - 1$). If particles do not overlap, the volume
filling factor is
\begin{equation}
  \frac{\sum_{i=0}^{N_{\rm p}-1} V_{{\rm cubes},i}}{V_{\rm j,1/2}} = 
\frac{\sum_{i=0}^{N_{\rm p}-1} 4\pi(2i+1) (\Delta r)^3}{\pi R_{\rm
    b}^2 \Delta r} =
    1 - \frac{1}{N_{\rm p}}\, .
\label{eq:Vfilling}
\end{equation}

Since we have a finite number of particles, the jet volume is only
partially patched by the volume occupied by such Lagrangian particles,
{\cblue i.e., the volume filling factor is smaller than one}.  {\cblue
  Increasing} the number of particles {\cblue brings it closer} to
one. To test the volume filling method, we produce an ``image'' of the
jet at an observer time $t_{\rm obs}=0$ with a $90^\circ$ viewing
angle, accumulating in each pixel the contributions corresponding to a
laboratory frame time interval $\Delta T=2 R_{\rm b}/c$. However,
instead of summing up the emissivity, we add up the length of the
intersection of each particle's volume with each pixel in the $(x_{\rm
  obs}, y_{\rm obs})$ plane (as described above). The idea behind the
substitution of the emissivity by the intersection length is that, at
$90^\circ$ the intersection length and the intersection volume of the
particles are proportional and, thus, measuring lengths or volumes is
equivalent.

Since we accumulate in every pixel all contributions in the range
$[-\Delta T/2, \Delta T/2]$, the intersection length with each
particle equals the size of the particle perpendicular to the LoS
($2\Delta r$). {\cblue Hence,} the value accumulated in a pixel ${\cal
  P}:=(x_{\rm obs}, y_{\rm obs})$, namely $L_{\rm px}$, {\cblue is}
\begin{equation}
L_{\rm px} = \sum_i \frac{A_i}{A_{\rm px}} 2\Delta r\, ,
\label{eq:Lpx}
\end{equation}
where $A_i$ and $A_{\rm px}$ are the area of intersection of a
particle with a pixel and the pixel area, respectively. The sum in
Eq.~(\ref{eq:Lpx}) extends over all particles that are intersected by
the line of sight that departs from ${\cal P}$. In the limit $\Delta r
\rightarrow 0$ (equivalently, $N_{\rm p} \rightarrow \infty$) $A_i
\rightarrow 4(\Delta r)^2$. On the other hand, the number of particles
intersected by the LoS departing from ${\cal P}$ and having a cross
sectional area $A_{\rm px}$ is $N_{\rm px}= A_{\rm px}/(2\Delta
r)^2$. Therefore, we have
\begin{equation}
  \lim_{\Delta r \rightarrow 0} L_{\rm px} = \lim_{\Delta r
    \rightarrow 0} 8 (\Delta r)^3 \frac{N_{\rm px}}{A_{\rm px}} = 2
 \sqrt{(R_{\rm b}^2 - y_{\rm obs}^2)}\, .
\label{eq:Lpx2}
\end{equation}
Equation~\ref{eq:Lpx2} simply expresses that, in the limit $N_{\rm p}
\rightarrow \infty$, the length measured in the pixel ${\cal P}$
should tend to the length of the chord determined by the intersection
of the jet body with the line of sight from ${\cal P}$.  {\cblue
  Figure~\ref{fig:circle}a shows that for $N_p\geq 16$ the results
  converge very rapidly} to the analytic expectation (thick black
line).
\begin{figure*}
\plotone{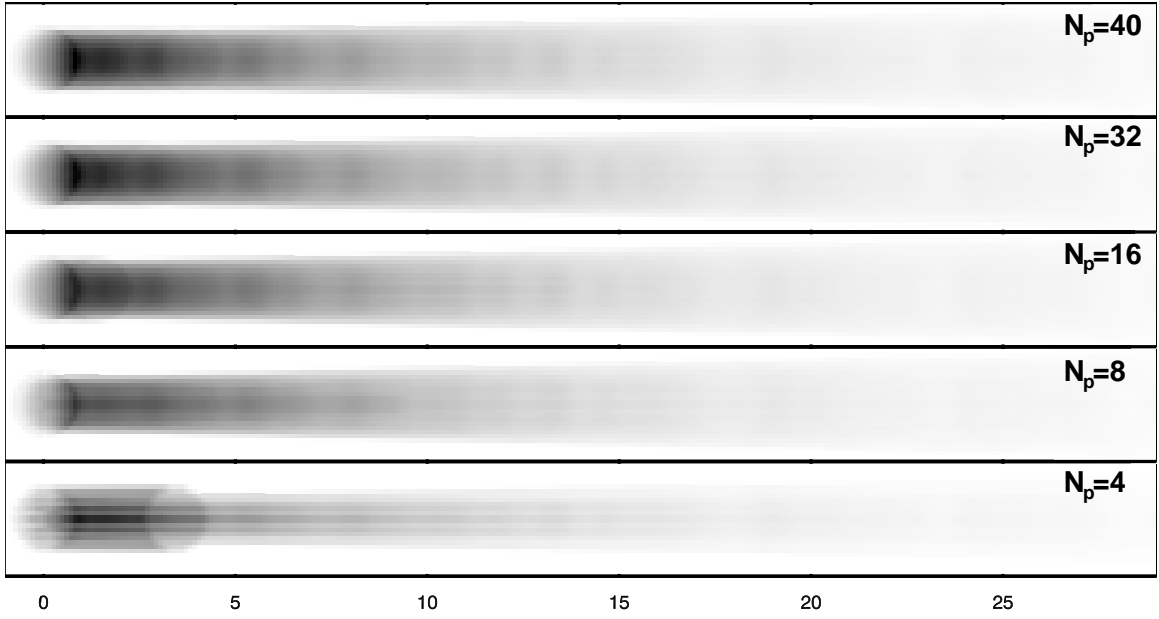}
\caption{Images of PM-L models used in the convergence test. From top to bottom $N_p = 40$, $32$, $16$, $8$ and $4$.
  \label{fig:PM-conv}}
\end{figure*}

\begin{figure*}
\plotone{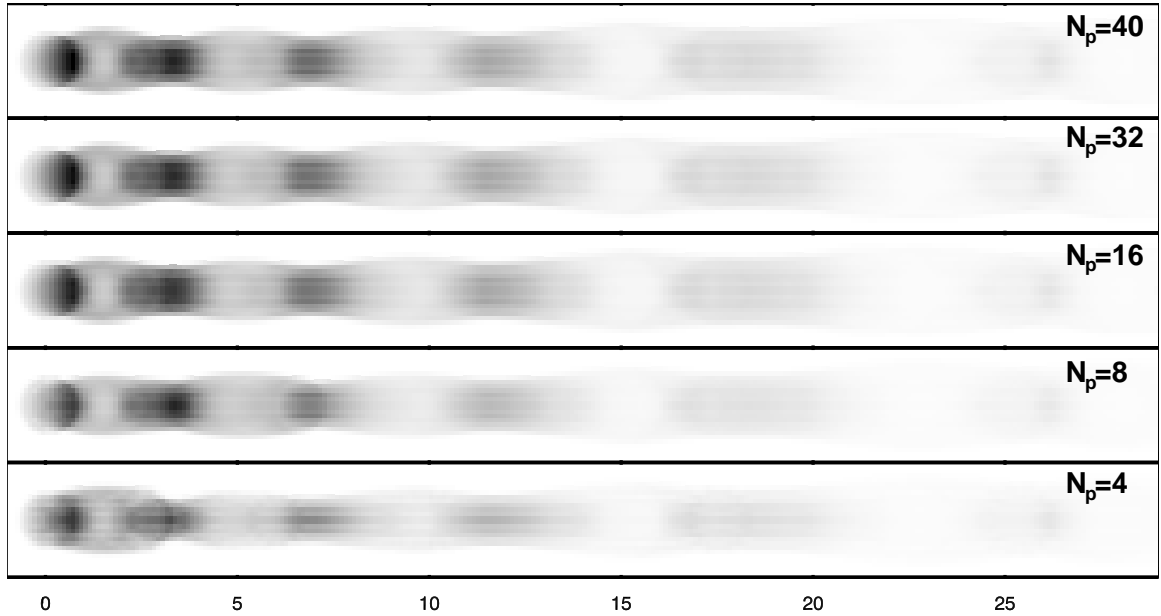}
\caption{Same as Fig.~\ref{fig:PM-conv}, but for the OP-L model.
  \label{fig:OP-conv}}
\end{figure*}

\subsubsection{Total flux}
\label{sec:totalflux}

To test the convergence of the imaging algorithm we have produced
images of quiescent {\cblue PM-L} and {\cblue OP-L} models with
varying $N_p$. {\cblue The total number of particles in the grid grows
  as $N_{\rm p}^2$, it is thus important to minimize the number of
  particle families for numerical purposes}. On Fig.~\ref{fig:circle}b
we show the total image flux at $15$\,GHz for {\cblue PM-L} and
{\cblue OP-L} models as a function of $N_{\rm p}$. The values are
normalized to the flux of the model with the largest number of
injected particles per beam radius ($N_{\rm p}=40$), which we consider
the reference value. This test is important because the total flux
represents a global value of every model, since it is computed by
summing up the individual fluxes arriving to each pixel in the
detector, and multiplying by the corresponding pixel area. Remarkably,
for $N_{\rm p}\geq 16$ the flux does not deviate more than $5\%$ form
the reference value. Thus, any model with $N_{\rm p}\geq 16$ has
sufficiently converged to an appropriate total flux. This has
motivated our choice to work with $N_{\rm p}=32$ in the current paper,
{\cblue since it yields an optimal trade-off between numerical
  accuracy and computational cost}.  Figures~\ref{fig:PM-conv} and
\ref{fig:OP-conv} show images corresponding to the convergence tests
for models {\cblue PM-L} and {\cblue OP-L}, respectively.

%

\end{document}